\documentclass[onecolumn,10pt,prl,letterpaper,groupedaddress]{revtex4}

\usepackage{times,amsmath,amsfonts,amssymb,latexsym,textcomp}
\usepackage{color}
\usepackage{graphicx}
\usepackage{multirow}
\usepackage{bbm}
\usepackage{tabularx}
\usepackage{amsthm}
\usepackage{dsfont}
\usepackage{placeins}
\usepackage[lofdepth,lotdepth]{subfig}
\usepackage[T1]{fontenc}
\usepackage[latin9]{inputenc}
\usepackage[english]{babel}
\usepackage{float}
\usepackage{array}
\usepackage{caption}
\usepackage{braket}
\usepackage{MnSymbol}
\usepackage{tikz}
\usepackage[electronic]{ifsym}
\usepackage{booktabs}
\usepackage{subfig}

\renewcommand{\phi}{\varphi}

\renewcommand{\epsilon}{\varepsilon}

\begin{document}

 \title{An Ontology of Nature with Local Causality, Parallel Lives, and Many Relative Worlds}
\author{Mordecai Waegell$^1$}
\affiliation{$^1$~Institute for Quantum Studies, Chapman University, Orange, CA, US}

\begin{abstract}
Parallel Lives (PL) is an ontological model of nature in which quantum mechanics and special relativity are unified in a single universe with a single space-time.  Point-like objects called lives are the only fundamental objects in this space-time, and they propagate at or below c, and interact with one another only locally at point-like events in space-time, very much like classical point particles.  Lives are not alive in any sense, nor do they possess consciousness or any agency to make decisions - they are simply point objects which encode memory at events in space-time.  The only causes and effects in the universe occur when lives meet locally, and thus the causal structure of interaction events in space-time is Lorentz invariant.  Each life traces a continuous world-line through space-time, and experiences its own relative world, fully defined by the outcomes of past events along its world-line (never superpositions), which are encoded in its external memory.  A quantum field comprises a continuum of lives throughout space-time, and familiar physical systems like particles each comprise a sub-continuum of the lives of the field.  Each life carries a hidden internal memory containing a local relative wavefunction, which is a local piece of a pure universal wavefunction, but it is the relative wavefunctions in the local memories throughout space-time which are physically real in PL, and not the universal wavefunction in configuration space.  Furthermore, while the universal wavefunction tracks the average behavior of the lives of a system, it fails to track their individual dynamics and trajectories.  There is always a preferred separable basis, and for an irreducible physical system, each orthogonal term in this basis is a different relative world - each containing some fraction of the lives of the system.  The relative wavefunctions in the lives' internal memories govern which lives of different systems can meet during future local interactions, and thereby enforce entanglement correlations - including Bell inequality violations.  These, and many other details, are explored here, but several aspects of this framework are not yet fleshed out, and work is ongoing.
\end{abstract}
\maketitle

\textbf{Overview}

This model has been developed in the spirit of Einstein \cite{einstein2015relativity}, Aharonov \cite{aharonov2008quantum}, and others, which can be loosely summarized by the sentence, \emph{``Once you can explain the workings of nature with the right physical story, the mathematical formalism will follow.''} I am not so presumptuous as to think that this is \emph{the right} physical story, but I do believe that this model possesses a remarkable degree of elegance and simplicity, and thus warrants serious consideration.  Here I attempt to present the most important parts of the story, after which I examine several specific details, examples, and questions.

The guiding principle for the development of this ontology is that there should be only one space-time in the universe, in which all objects reside, and the only fundamental objects should be point-like `lives' that move along world lines like classical point particles, interacting only with one another at events in space-time, so that local causality is explicitly obeyed by all contents of the universe.  All higher-order physical phenomena in the universe must be characterized in terms of the lives they comprise.  This guiding principle is motivated by the fact that obtaining a quantum theory in 3-space, which strictly obeys local causality, would automatically clear out all of the conceptual difficulties associated with configuration space, nonlocality, and causal structure, and could lend itself much more naturally to unification with general relativity.  The PL mechanism for treating entanglement seems to enable this type of quantum theory, and thus I feel it will be valuable to develop it to its logical conclusion.

Lives in the PL model \cite{brassard2013can, waegell2016locally, brassard2017parallel, brassard2017equivalence}\footnote{The model presented here has grown outside of the other versions, and the distinction between `lives' and `relative worlds' is new.} are the general point-like elements that collectively form all of the generalized quantum fields in the universe, and each life is an infinitesimal grain of the field, and carries all of the field's local physical properties.  Lives of the quantum fields are never created nor destroyed, and each one follows a single continuous world-line through space-time.  The use of the word `lives' in PL has nothing to do with any notion of biological life, consciousness, or decision making --- they are simply point objects that move on a world-line and encode a local memory of events.  A life's experience is defined by the specific list of outcomes (never superpositions) at each interaction event along its past world-line, which are stored in the \emph{external memory} of the life.  The different outcomes of an interaction correspond to different relative worlds the life of a given system may experience --- and this list of outcomes is in the preferred basis, such that no life experiences being in a superposition state.  Each life also possesses an \emph{internal memory}, which also updates at local interaction events, but which is not a direct part of the life's experience.  The internal memory carries a local \emph {relative wavefunction}, which is a quantum state in the Hilbert space of all systems in the \emph{past interaction cone} of the given life, defined by the list of initial states of each system (before the local coupling), and a list of pairwise unitary couplings, between all systems within its past interaction cone.  In other words, each life is a point-like information-gathering-and using-system (IGUS) \cite{zurek1990algorithmic, atmanspacher2002determinism}, although perhaps it is more appropriate to think of them simply as information-gathering systems, and nature itself as the mechanism that uses this local information to determine local dynamics.  I will make this analogy explicit in the appendix by providing a classroom exercise in which students play the roles of lives of different systems and carry out a Bell experiment.

To summarize very explicitly, the axioms of this model are: 1) The only fundamental objects that exist in nature are point-like lives which move along world-lines through relativistic space-time.  2) At local interaction events, lives encode internal memories which contain local relative wavefunctions.  There is no fundamental universal wavefunction.  3) At local interaction events, lives experience definite outcomes in a specific preferred basis --- never superpositions --- and encode those outcomes into external memory.

The main goal of this article is to demonstrate that it is possible to find a model of nature which begins with these axioms and reproduces all of observed quantum mechanics --- including violation of Bell inequalities --- without any action-at-a-distance.

Each excitation of the quantum field, such as a particle, is characterized by the coordinated motion of a specific subset of lives of the field --- noting that not all excitations need to fit the particle model in PL.  All physical fields in space-time are composed of lives, and thus all physical fields are relativistically retarded.  The subset of lives belonging to a given relative world of a system interact with one another locally, giving rise to the usual wave-like dynamics, but they are always invisible to one another.  In contrast, an interaction between two different physical systems comprises a specific set of point-like one-to-one interaction events in which each life of one system experiences a meeting with a life of the other system.

For an isolated physical system, the density of lives $P(x,t)$ at a particular point in space-time is associated to the probability density $|\psi(x,t)|^2$ (or its relativistic equivalent) in quantum mechanics.  In general, $P(x,t)$ of each individual system is associated to its reduced density matrix in quantum theory, and the dynamical equations for $P(x,t)$ are local.  Note also that the continuity of world-lines for the parallel lives is equivalent to the conservation of probability in quantum theory, which is also to say that the total number of lives in the universe is constant.  The entire model is consistent at this level with a pure universal wavefunction description --- but the wavefunction does not tell the whole story, because it describes only the collective unitary dynamics of the lives, while remaining agnostic to their individual dynamics.  It is the local dynamical equations of lives in space-time which are fundamental in PL, and not the Schr\"{o}dinger or Dirac equations.  The individual dynamics of the lives are necessary to correctly describe the different relative worlds in PL, and the local mechanism which produces entanglement correlations.  Where the universal wavefunction fails to provide a specific ontology in space-time, the PL model does so by replacing the universal wavefunction with an infinite collection of local relative wavefunctions, which are fully defined by their past interaction cones.  This may seem a highly redundant mechanism, but consider that the universal wavefunction seems to act everywhere in space-time at once, and thus the entire thing is physically located at every point in space-time, and not just the portion belonging to the past interaction cone, which actually seems more redundant.  Einstein and Lorentz discussed this very problem with the universal wavefunction in configuration space with Schr\"{o}dinger \cite{przibram1967letters, norsen2017foundations}, and Schr\"{o}dinger made some attempts to put the theory back into 3-space using reduced density matrices, but he soon abandoned this.  This idea was revived in \cite{allori2011many}, and PL develops it into a fully local theory in space-time.

While PL is consistent with some of Everett's ideas \cite{everett1957relative, everett1973theory}, unlike most many-worlds models \cite{dewitt2015many, wallace2002worlds, wallace2012emergent, wallace2010decoherence, wallace2003everett, saunders2010many, vaidman2002many, sebens2014self}, there are no objective (or global) \emph{worlds} that exist throughout the single objective space-time of this model.  Instead, there are many \emph{relative worlds}, fully defined for each individual life by the complete local interaction history along its own past world-line and past interaction cone --- and this is the only definition of a \emph{world} in PL.  When a physical system undergoes a collapse into one of several outcomes during an interaction event, the lives of the system subdivide locally into different relative worlds, where each relative world corresponds to an orthogonal term in the wavefunction, in a specific preferred basis.  This division of lives has absolutely no effect on any other physical system that did not participate in the local interaction event.  To be explicit, unlike some other many-worlds interpretations, there is no objective world that exists throughout space-time, and which then splits (or bifurcates) instantaneously into multiple objective worlds that exist throughout space-time --- worlds are only defined by world-lines.  I suspect that this perspectival \emph{many-world-lines} model is closer in spirit to Everett's notion of \emph{relative states} than some other many-worlds models.  I should also note some other works \cite{allori2011many, deutsch2000information,timpson2005nonlocality, wallace2010quantum, brown2014bell, albert1988interpreting} which seem to share some specific features or ideas in common with the model presented here.

The relative wavefunctions, $\psi_i$, carried in the internal memories of each point-like life $i$ of each system also enable entanglement correlations to be obeyed using the explicitly local PL mechanism.  This is possible because whenever lives of different systems meet at an interaction event, the relative wavefunctions they each carry become synchronized, since they now share the same past interaction cone.

The relative wavefunction of a given life can be seen as just some portion of a pure universal wavefunction $\Psi$ (defined in some specific Lorentz frame) --- but crucially, each portion has been accumulated locally by a specific point-like object in space-time as it underwent local interactions on its past world-line.  In this way, all of the information comprised by the entire universal wavefunction is physically distributed among the memories of lives throughout space-time --- each fully defined by the chain of interactions in its past light cone.  The universal wavefunction $\Psi$ in configuration space is merely an objective bookkeeping tool, and not a fundamental part of the PL ontology --- unlike most many-worlds, Bohmian \cite{bohm1952suggested, holland1995quantum, wyatt2006quantum, hiley2001non}, and Consistent Histories \cite{griffiths2003consistent,gell2012decoherent} interpretations of quantum mechanics.  The relative wavefunctions $\psi_i$ may contain internal structure defined in a configuration space, but each $\psi_i$ is assigned to a single specific point in space-time.  As a result, the $\{\psi_i\}$ of PL are a Lorentz invariant description --- meaning that the set of all $\{\psi_i\}$ can be thought of as a relativistically retarded field.

Now I will begin to explore the dynamics of lives in space-time.  For interference effects in an isolated quantum system, such as the 2-Slit or Mach-Zehnder interferometer experiments, the description is much the same as in other many-worlds models, with the added detail that in PL, each life carries its own relative wavefunction hidden in its internal memory, whose phase accumulates as that life propagates along its trajectory through space-time.  When multiple lives of a system in the same relative world meet in the same location, their collective relative wavefunctions govern their collective instantaneous dynamics, producing the usual interference effects.  In this way, the different lives within the same relative world interact with one another, very much like the \emph{many-interacting-worlds} \cite{hall2014quantum} or hydrodynamical \cite{madelung1926hydrodynamical, trahan2005multidimensional, schiff2012communication} models.

For the Mach-Zehnder interferometer, half of the lives of the particle reflect at the first beam splitter, and half transmit.  Note that this does not separate these lives into different relative worlds, since they do not become entangled with the beam splitter\footnote{More on the single-system approximation later.}.  The lives on each arm accumulate phase as they move along their respective trajectories, and when they meet at the second beam splitter, these local phases create the expected interference.  If one arm of the interferometer is obstructed, as in the Elitzur-Vaidman experiment \cite{elitzur1993quantum}, then that half of the lives never arrives at the second beam splitter, and the interference does not occur.

In order to understand the local mechanism which governs entanglement correlations in PL, we must make the paradigm shift to the physical interpretation of the wavefunction of an entangled state in terms of internal memories.  Consider the classic entanglement example of the Einstein-Podolsky-Rosen (EPR \cite{einstein1935can}) state (or Bell state) of two spins, $|\psi\rangle = (|00\rangle + |11\rangle)/\sqrt{2}$.  It has become conventional to imagine that there is some nonlocal connection between the two spins, and that they somehow act collectively as a single delocalized object --- even if that object obeys the No-Signalling principle.  In PL, there can never be nonlocal correlations of any kind, and this is all governed by internal and external memories of lives.  The reduced density matrices always determine the proportion of lives of each system that experience each relative world, and the proportionalities that determine how the lives of each spin separate into different relative worlds if they are measured, are also determined by its reduced density matrix and the Born rule.  The set of correlation rules in $|\psi\rangle$ are carried locally in the internal memory of each of the lives of both spins, as if each life carries a hidden classical notebook with ``$|\psi\rangle$'' written inside, and nature reads and writes information between the notebooks of different lives when they meet at local interaction events.

In this scenario and many others, it is a fine approximation to think of macroscopic systems like Alice and Bob as having their own collective lives, even though this is not the true microscopic story.

Suppose that one spin is sent to Alice and the other to Bob, who both measure the $\{|0\rangle, |1\rangle \}$ basis of their spin.  Using the reduced density matrix we obtain probability $50\%$ for each outcome of either measurement, and thus the lives of Alice and her spin divide evenly into relative worlds with $|0\rangle$ and $|1\rangle$ respectively, and likewise for Bob and his spin.  The hidden information $|\psi\rangle$ is passed along from spin to measurer during this local interaction. Note that the entanglement correlations have not yet been obeyed, as they would have to be if Alice and Bob coexisted in a single objective (global) world --- as long as they are space-like separated, Alice's relative worlds are completely independent from Bob's relative worlds.

Now, Alice and Bob meet locally to compare the results of their measurements, and the hidden information $|\psi\rangle$ carried locally in the internal memories of all of their respective lives governs the meeting between Alice's and Bob's lives as follows:  The $50\%$ of Alice's lives in a $|0\rangle$ world only meet the $50\%$ of Bob's lives in a $|0\rangle$ world, and the $50\%$ of Alice's lives in a $|1\rangle$ world only meet the $50\%$ of Bob's lives in a $|1\rangle$ world.  As seen by all lives of Alice and Bob, the entanglement correlations of $|\psi\rangle$ have now been obeyed, without the need for superluminal influences of any kind --- and a Bell inequality \cite{bell1966problem, bell2001theory, bell2004speakable} can be violated as usual.

According to their individual interaction histories, some lives of different systems are unable to interact, and are thus invisible to one another.  While the rule governing which lives can meet and which are respectively invisible may seem somewhat \emph{ad hoc}, it is explicitly necessary in order to prevent lives from experiencing routine violations of physical conservation laws.  For example, if an experimenter measures the energy of a system, and both the experimenter and the system split into two relative worlds corresponding to two different energy eigenvalues, then the lives of the experimenter who saw one energy must be invisible to the lives of the system with the other energy, and vice versa --- otherwise if the experimenter measures the energy again, they may find that the energy had spontaneously changed to the other eigenvalue.  At the fundamental level, all entanglement correlations arise as superpositions of different states that all obey the same conservation equation.  As a result, we can think of the entanglement correlation rules produced by the relative wavefunctions in the internal memories of all lives as a local mechanism for enforcing the fundamental conservation laws of nature.

\begin{figure}
\begin{center}
\caption{Minkowski diagram showing the world line (blue) and past interaction cone (green) for a particular life as it propagates through space-time.  Interaction events are shown as dots.  The world lines of other lives outside the past interaction cone are shown in grey, and are outside the relative world of the (blue) life.}
\includegraphics[width=4in]{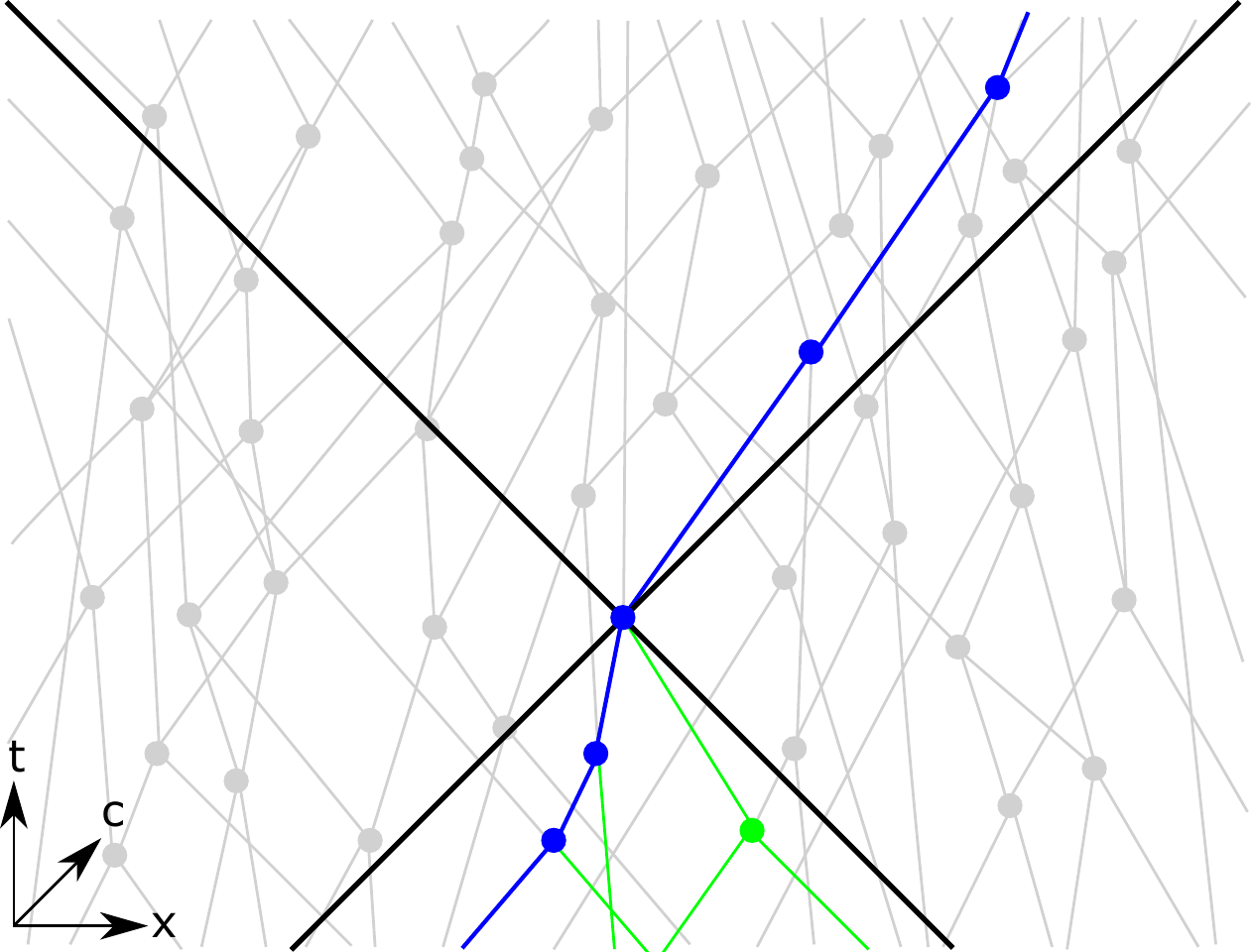}\label{IntCon}
\end{center}
\end{figure}

The local mechanism for handling entanglement correlations using the relative wavefunctions carried by all lives, generalizes to all interactions in the universe.   This shows precisely how all of the information comprising the universal wavefunction $\Psi$ of configuration space is physically distributed among the $\{\psi_i\}$ in the internal memories of lives throughout space-time, and propagates from event to event, always obeying local causality.  Specifically, the complete set of interaction rules are:
\begin{itemize}
\item Each relative world experienced by a life is completely defined by a discrete list of specific outcomes for all systems in every pairwise interaction within its past interaction cone (see Fig. \ref{IntCon}).  If two lives of different systems have different outcomes for the same event in their external memories, then those two lives are invisible to one another and cannot meet (and if they did, then afterwards it would be impossible to define consistent relative worlds for either system).  This is also required in order to guarantee that all lives experience the obedience of all conservation laws in nature.  Each relative world is like a pseudo-classical history, and we will index them by $h$.  For each relative world $h$, we define the history state $|\xi^h\rangle$ as the tensor product of the most recent outcome states of every individual system in the past interaction cone, noting that different relative worlds can still have the same history state.
\item When lives of two different systems meet at an interaction event in space-time, the relative wavefunctions in their internal memories become synchronized.  Specifically, if the life of the system 1 had interactions $\{U_i\}$ and initial states $\{|\psi\rangle^k  \}$ of systems in its past interaction cone, and the life of system 2 had interactions $\{U_j\}$ and initial states $\{|\psi\rangle^l  \}$ of systems in its past interaction cone, then after this interaction event, they now both have $\{U_i, U_j\}$ and initial states $\{|\psi\rangle^k , |\psi\rangle^l \}$ in their internal memories, and their relative wavefunction $|\psi\rangle$ is the state obtained by applying all of the coupling unitaries to all of the initial states in the order given by the past interaction cone.
\item There is a new local coupling unitary $U^{1,2}$ between the two interacting systems, which is now added to their internal memories, and the relative wavefunction $|\psi\rangle$ is updated accordingly.  The preferred basis of both systems involved in the interaction may also change as part of the interaction.
\item In order to determine how the lives within each of the existing relative worlds are distributed into the new relative worlds (the new preferred basis), we first consider how the coupling unitary $U^{1,2}$ acts on the history state $|\xi^h\rangle^{1,2}$, $U^{1,2}|\xi^h\rangle^{1,2}|\xi^h\rangle^{O}, = \sum_{i,j} a_{ij} |u_i\rangle^1 |v_j\rangle^2|\xi^h\rangle^{O}$, where $|u_i\rangle^1$ and $|v_j\rangle^2$ are the outcomes each system experiences in the new preferred basis, and $|\xi\rangle$ has been factored into two groups, one for systems 1 and 2, and another for all other systems in the past interaction cone, which we label collectively as $O$.  Lives from the relative world $h$ experience the new outcome $|u_i\rangle^1 |v_j\rangle^2$ if and only if $a_{ij} \neq 0$, and this defines the set of possible futures $f^h \in \{(i,j)\}$ of $h$.  Next, the fraction of the lives from history $h$ that experience each future $(i,j)$ in  $f^h$ is computed as $w^h_{ij} = \frac{\big|\langle \psi|^{1,2,O}|u_i\rangle^1 |v_j\rangle^2|\xi^h\rangle^{O}\big|^2}{\sum_{(i',j') \in f^h} \big|\langle \psi|^{1,2,O}|u_{i'}\rangle^1 |v_{j'}\rangle^2|\xi^h\rangle^{O}\big|^2}$, and we also set $w^h_{ij} = 0$ for $(i,j) \not\in f^h$.  Finally, the total proportion of lives with history $h$ and future $(i,j)$ is given by $P^h_{ij} = w^h_{ij}P^h$, and the outcome of the meeting is recorded into their external memories.
\item I should stress that this rule makes an arbitrary assumption about the trajectories of individual lives, and is only intended as a proof of principle, showing that the overall proportionality currents can be correctly conserved.  When two systems interact, it is not yet clear how the lives which share the same relative history will distribute among their relative futures, since I do not yet have a model of the dynamics and trajectories of individual lives.
\end{itemize}

While the reduced density matrices of the individual systems do not appear explicitly in this process, it can be easily verified that for a given system $A$, the total proportion of lives which experienced outcome $|u_i\rangle^A$ (regardless of their histories) during its most recent interaction is given by $P(|u_i\rangle^A) = \langle u_i|^A \rho^A | u_i \rangle^A$, where $\rho_A$ is the reduced density matrix of system $A$ alone.  Likewise, if the system interacts again, and the preferred basis is changed to $|u'_j\rangle^A$, the total proportion of lives (again, ignoring different histories) that experience each outcome is $\langle u'_j|^A \rho^A | u'_j \rangle^A$.  Thus, the physical distributions of the lives of different systems in space-time are given by the reduced density matrices of those systems.  At this level, this model is consistent with \cite{przibram1967letters, allori2011many, norsen2017foundations}, with the added refinement that the relative wavefunctions enforce entanglement correlations by governing the proportions with which lives of different systems meet during interaction events.

Within the relative worlds of PL, reality itself propagates through local interactions at or below the speed of light, with experienced reality defined in each relative world by the complete history of local interactions within the past interaction cone.  A remote event has no reality for a given life until a signal from that event arrives and interacts with it.  I call this \emph{quantum relativity}, and it explains how PL resolves Wigner's-friend-type scenarios \cite{wigner1995remarks}, including the Bell state example; the lives of the cat separate into different relative worlds when they interact with the poison apparatus, and then the lives of Schr\"{o}dinger separate into different relative worlds when he opens the box, and then the lives of Wigner separate into different relative worlds when he open the door of the laboratory, and so on --- the division of relative worlds always occurs at a local interaction between two systems.  \footnote{Noting again that in most classical situations, we can treat macroscopic systems as having their own collective lives, even though this is only approximation to the true microscopic story}.  When Alice makes her measurement, her lives in the $|0\rangle$ and $|1\rangle$ worlds have absolutely no information about Bob's present state, which means that Bob's current remote state is excluded from Alice's reality, and vice versa.  Alice's relative worlds only acquire information about Bob when their world-lines intersect at a local interaction event, and only then is one of Bob's outcomes realized for the worlds of Alice, and vice versa.  As the world lines separate, the mutual exclusion is immediately restored, but the relative worlds of Alice and Bob now include the correlated outcomes of that event in their mutual histories --- which is the only definition of physical correlation in PL --- and they now carry hidden relative wavefunctions that guarantee that if they interact again in the future, the results will be consistent with the correlations at the previous interaction (i.e., within each relative world, consistent with a collapse having occurred at the previous interaction).

In PL, systems that are already entangled in a universal wavefunction representation may still be physically uncorrelated until they locally interact, or exchange signals that obey local causality.  By making this paradigm shift in how we interpret the wavefunction of quantum mechanics, we regain an interpretation of nature that strictly obeys local causality, and whose only contents are point-like objects that follow world-lines through Minkowski space \cite{minkowski2013space}.  The refined PL picture of many relative worlds also specifies exactly where in space-time collapse events are experienced by each life, and the causal structure of events in space-time is Lorentz invariant.

In contrast, explaining Bell correlations in an objective (global) world (one or many) that exists throughout space-time using superluminal or atemporal influences between Alice and Bob --- even if they obey the No-Signalling principle --- still requires the injection of causal information that originates {\it outside} of the causal structure of space-time.  Furthermore, in different Lorentz frames, this information may appear to be first injected into space-time at different point-like events, and thus the causal structure of events and collapses becomes highly nontrivial.  The causal structure for an objective world can be preserved by choosing a preferred Lorentz frame for nature and allowing faster-than-light influences, but this seems especially at odds with the symmetry of special relativity.\\

\textbf{Conceptual Details}

In the nonrelativistic limit of PL, the collective configuration of lives in a physical system evolves according to the Schr\"{o}dinger equation, insofar as PL agrees with its empirical predictions.  When the average interactions with other systems are weak enough to be approximately characterized by a potential $V(x)$, this gives the usual unitary dynamics without wavefunction collapse.  When the interaction between two systems is stronger, the Hamiltonian must instead include an explicit coupling term that creates entanglement between them, and as the relative wavefunctions divide into orthogonal terms, the lives of the two systems divide into separate relative worlds.  I call cases where $V(x)$ is a valid description the \emph{single-system approximation}.

In the (unkown) exact dynamical theory, the only potentials in the Hamiltonian are local interactions between lives that meet at events in space-time.  For example, when considering scattering between two charged particles, one must consider the local interactions between lives of the particles and also the lives of the quantum electromagnetic field that propagate through the space between them.  In the detailed picture of PL, the coupling between the two particles happens through a combination of local propagation of lives and local interactions between them.

The configuration of the lives of a system evolves according to $\rho(x,v,t)$, which is the density of lives at position $x$, with velocity $v$ ($|v|\leq c$), at time $t$.  Integrating out the velocities gives the average collective motion of the lives, $P(x,t) \equiv \int_{-c}^{c}\rho(x,v,t)dv = |\psi(x,t)|^2$ (or the relativistic equivalent of $|\psi(x,t)|^2$), and the probability current density is given by $J(x,t) \equiv \int_{-c}^{c}\rho(x,v,t)v dv$.   In nonrelativistic situations where the $V(x)$ approximation holds, this density evolves according to the single-system Schr\"{o}dinger equation.  While $\rho(x,v,t)$ is a joint proportionality distribution for $x$ and $v$, it should not be confused with a Wigner function, which concerns the joint probability distribution for outcomes of different possible measurement interactions between two systems --- where the single-system approximation no longer applies.  In general, the probability to measure the system with momentum $p$ is not given by the marginal over the positions, i.e., $P(p,t) \neq m\int_{-\infty}^{\infty}\rho'(x,p,t)dx$, where the specific relation between $\rho'(x,p,t)$ and $\rho(x,v,t)$ depends on the details of the system.

There is a physically preferred basis for the entire Hilbert space of the universal wavefunction, and the different relative worlds are defined as the eigenstates in this basis.  The preferred basis for a given system is determined by the details of its most recent interaction event.  For example, the lives of a spin that is measured in the $\sigma_z$ basis divide into two relative worlds corresponding to the two eigenstates of $\sigma_z$, and thus the $\sigma_z$ basis is the preferred basis after the measurement.  Furthermore, there is always a preferred separable basis for the systems in an entangled state, and the lives of each system are divided into relative worlds in this basis.  For example, a Bell state $(|0\rangle|0\rangle + |1\rangle|1\rangle)/\sqrt{2}$ could have the preferred basis $\{|0\rangle, |1\rangle\}$ for each qubit, meaning these are the relative worlds its lives experience.  On the other hand, the same state could just as well have the preferred basis $\{(|0\rangle + |1\rangle)/\sqrt{2}, (|0\rangle - |1\rangle)/\sqrt{2} \}$ for each qubit, and the lives would experience different relative worlds, which is why the preferred basis must be specified in PL --- every life must have a single definite experience everywhere on its world-line.

\begin{figure}[h]
\includegraphics[width=6in]{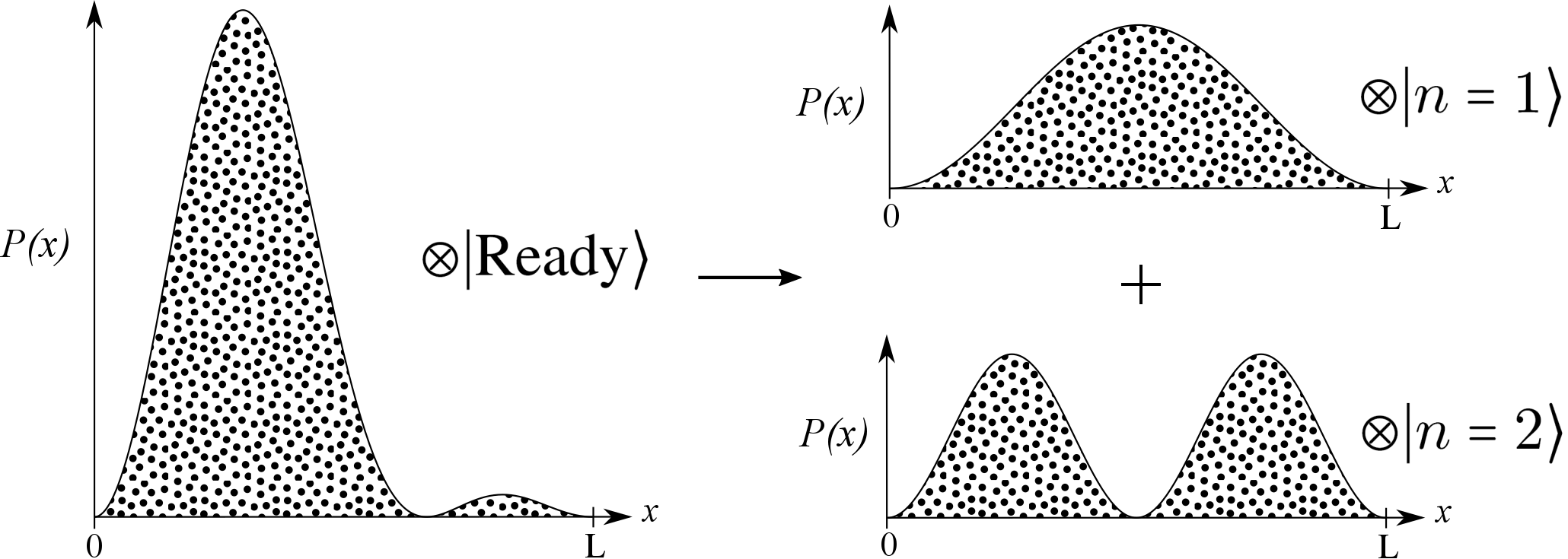}
\caption{An experimenter, initially in state $|\textrm{Ready}\rangle$, measures the energy $E_n$ of an electron trapped in an infinite square well.  The initial separable state of the experimenter and electron is $|\psi\rangle = \big[ \sin\big(\pi x / L  \big) + \sin\big(2 \pi  x / L  \big)   \big] |\textrm{Ready}\rangle$.  After the interaction, the experimenter and electron are entangled in the state $|\psi'\rangle =  \sin\big(\pi x / L  \big)|n=1\rangle + \sin\big(2 \pi  x / L  \big)    |n=2\rangle$, and both systems have divided into two relative worlds in this preferred basis.  The total probability density in 3-space of the lives of the electron must redistribute from $P(x) = \big[ \sin\big(\pi x / L  \big) + \sin\big(2 \pi  x / L  \big)   \big]^2$ to $P'(x) = \sin^2\big(\pi x / L  \big) + \sin^2\big(2 \pi  x / L  \big)$, where the latter is obtained by tracing the experimenter out of $|\psi'\rangle\langle \psi'|$.  Because individual lives cannot move faster than $c$, this redistribution cannot be instantaneous, and thus there is a speed limit on the experience of collapsing into either of these two relative worlds.  The actual trajectories and relativistic dynamics of individual lives are not yet known, so it is not clear which half of the lives of the electron in distribution $P(x)$ become part of the $\sin^2\big(\pi x / L  \big)$ distribution and which half become part of the $\sin^2\big(2\pi x / L  \big)$ distribution.  Regardless, the half in $\sin^2\big(\pi x / L  \big)$ become invisible to the half of the lives of the experimenter that got $|n=2\rangle$, and vice versa, and likewise for the halves in $\sin^2\big(2 \pi x / L  \big)$ and $|n=1\rangle$.}

\label{SquareWell}
\end{figure}

As a result, the dynamics of measurement are explicitly context-dependent, since as the wavefunction transitions from the initial state $\psi_0(x) |R\rangle = \sum_i a_i \psi_i(x)|R\rangle$, to the entangled state $\sum_i a_i \psi_i(x)|M_i\rangle$ during a measurement interaction, the lives of the system must redistribute themselves from $P_0(x) =|\sum_i a_i \psi_i(x)|^2$ to  $P_1(x) =\sum_i |a_i \psi_i(x)|^2$ --- where $\psi_i(x)$ are the eigenstates (context) of the observables that were measured (which is now the preferred basis), $a_i$ is a normalized set of coefficients, and $|R\rangle$ and $|M_i\rangle$ are the `ready' and `indicator' states of the measurement device.  Note that the cross-terms in $P_0(x)$ are gone after the collapse into the different worlds in $P_1(x)$ has finished, as shown in the example of Fig. \ref{SquareWell}.

Because individual lives are constrained by the speed limit $c$, the separation of the lives into different relative worlds during an interaction cannot generally be instantaneous --- it takes a finite time to complete, as determined by the initial state and measurement basis.  This aspect of PL is a strict departure from the instantaneous collapse of the Copenhagen model, obtained by imposing a new type of relativistic constraint, which may suggest new experiments to probe the dynamical nature of collapse and the splitting of worlds --- and possibly to falsify PL as a physical theory.

For measurements of the position, the probability density is literally given by the existing density of lives in each position, and no motion is required to divide the lives into position eigenstates.  As a result, this splitting of relative worlds may be genuinely instantaneous.  Observables that do not commute have the characteristic property that the `collapse' process from eigenstates of one to eigenstates of the other cannot be instantaneous, because it takes time for the lives to redistribute.

The proportion of lives that experience each outcome of a measurement is fundamentally noncontextual, since it is fully determined, as usual, by the modulus-squared inner product of the initial and final states.  However, the Kochen-Specker theorem \cite{bell1966problem, kochen1967problem, spekkens2005contextuality} remains valid for the experience of individual lives, and we do not yet know their explicit dynamics.  The position of each life acts as a noncontextual hidden variable, since this pre-existing property is revealed by a measurement of the position, regardless of which other commuting observables are measured.  To avoid confusion, I should stress that even though each life always has a definite location in space-time, that location is not contained in its external memory (e.g. it is hidden in internal memory, but is not part of the experience of that life) unless it was involved in an interaction with another system wherein relative worlds were separated in the position basis --- and then it only remembers the location it had at that one event.  If the momentum is measured, then afterwards the lives remember their momenta, but not their positions, since the lives with each momentum eigenvalue have spread out into the shape of the corresponding momentum eigenstate --- in agreement with the Heisenberg uncertainty principle.  If the local dynamics of PL is found to be deterministic, then there must also be some mix of contextual and noncontextual hidden variables that predefine the outcome each life will experience for all measurements of all other observables.  If the dynamics is found to be nondeterministic, then some or all of these other outcomes may be randomly determined during the interaction process.

It is important to stress that regardless of determinism, the relative world a life experiences is fully characterized by the outcomes of a particular set of mutually commuting observables during its last local interaction.  Noncommuting observables that had definite values in the previous relative worlds may no longer have definite values, and the same is true of future relative worlds if a noncommuting observable is measured next.  The reality experienced by lives in PL is composed from a series of past interaction outcomes, and there is no underlying ontology in which all observable properties of a system are simultaneously defined.  Depending on the question of determinism, this may not be the case for individual lives, which may, in principle, possess definite values for all observable properties, even if those values are not written in their external memories.

A quantum system comprises a countably infinite number of parallel lives, where the countable infinity belongs to the Nonstandard Natural Numbers \cite{robinson2016non, albeverio1986nonstandard}, which contain a monotonic succession of distinct infinities, such that the ratio of two such countable infinities is a finite rational fraction.  When the lives of a system divide into different relative worlds, the number of lives in each world is still countably infinite, but in general these are infinities of different sizes, whose relative proportions are determined from the wavefunction by the Born Rule (approximating irrational probabilities with rational fractions to infinite precision).  The world-line traced by each life defines a possible experience of the universe.  The Born Rule probabilities in ensemble statistics emerge directly from the relative proportion of lives that go into each relative world, because the individual lives of a given system are invisible to one another, and have no control over which relative world they will end up in.

The experience of the classical world emerges from the microscopic PL description in large composite systems that exchange relatively frequent internal interactions.  Using the rules of the local entanglement mechanism, these frequent interactions quickly disseminate any hidden entanglement correlation information throughout the entire composite system. This causes macroscopic objects to collapse relative to a classical observer at the speed with which this information propagates through the environment --- something physically akin to a speed of sound, which will actually be closer to $c$ in most classical environments, and equal to $c$ for some fields.  The frequent interaction waves passing from one side of a macroscopic object to the other are what enable the approximation that the macroscopic object has collective lives of its own, which I have used in numerous examples.  There is no fundamental new physics that emerges at the classical scale --- the effective collapse (the separation of lives into different relative worlds) at a microscopic event spreads locally from interaction to interaction until lives of a macroscopic number of systems in the environment have all separated.   This is similar to, but somewhat different than, various models of decoherence \cite{joos2013decoherence, zurek2003decoherence, schlosshauer2005decoherence, zurek2006decoherence}, which make different arguments about the origin of the preferred basis in classical environments.

 In the example of Fig. \ref{SquareWell}, an experimenter has an electron trapped in an infinite square well and isolated from the environment, in an equal superposition of the ground and first excited states, and then measures the energy of the electron.  The initial product state of the electron and measurement apparatus is $(|1\rangle + |2\rangle)|R\rangle/\sqrt{2}$, which evolves into the entangled state $(|1\rangle|M_1\rangle + |2\rangle|M_2\rangle)/\sqrt{2}$, where $|R\rangle$, $|M_1\rangle$, and $|M_2\rangle$, are the ready state and indicator states of the measurement device.  The role of the measurement device is to amplify the outcome of this quantum measurement to the classical regime.  This means that the measuring device rapidly interacts with neighboring systems of the environment, and in a relatively brief amount of time, the hidden entanglement rules propagate through the entire laboratory environment, resulting in the entangled state $(|1\rangle|M_1\rangle|E_1\rangle+ |2\rangle|M_2\rangle|E_2\rangle)/\sqrt{2}$ --- where $|E_1\rangle$, and $|E_2\rangle$ are orthogonal states of the environment corresponding to the two outcomes.  Likewise, as the wave propagates, the lives of the electron, the measuring device, and the entire classical environment divide into separate relative worlds --- corresponding to all of these systems becoming mutually entangled in the universal wavefunction.  Note that environmental decoherence can cause exactly the same entanglement to arise, even without the measurement taking place.

The entanglement correlations between the lives of the electron and the experimenter work no differently than in the Bell state example above:  If the experimenter measures the energy of the electron again, then the lives of the experimenter in relative world $|E_1\rangle$ only meet the lives of the electron in relative world $|1\rangle$, and likewise for the lives $|E_2\rangle$ and $|2\rangle$, and thus the experience of each life of the experimenter is consistent with a collapse having occurred during the first measurement.

PL solves the quantum measurement problem for entangled states using local causality, but the local mechanism of propagating hidden information in different relative worlds is actually more general than quantum mechanics, and allows for a description of any Non-Signalling correlations, up to any including the Popescu-Rohrlich (PR) Box \cite{popescu1994quantum}.  Constraints like Tsirelson's bound \cite{cirel1980quantum}, Heisenberg uncertainty relations \cite{heisenberg1966uncertainty} between noncommuting observables, and contextuality, originate in the algebraic structure of quantum mechanics, and it is the union of the PL mechanism and quantum theory that gives a complete physical model.

Clearly not all of the details of this model have been worked out --- my main goal here is to present a coherent and comprehensible framework in which quantum physics may be explained consistently in a single objective Minkowski space-time occupied by point-like objects, with a Lorentz-invariant causal structure.  Individual lives must propagate at or below $c$, and we know the collective dynamics from many solutions to Schr\"{o}dinger and Dirac equations, in both the relativistic and non-relativistic limits, but the general relativistic dynamics of the individual lives remain unknown.  It is not that viable solutions do not exist, since it seems the relativistic Bohmian trajectories could suffice, but rather that the problem seems underconstrained within the present framework, leaving many possibilities open.  I may still be missing a key insight that will pin things down. \\

%%%%%%%%%%%%%%%%%%%%%%%%%%%%%%%%%%%%%%%%%%%%%%%

\textbf{The Local Entanglement Mechanism}

To ensure the description is clear, I now consider several examples in detail.  In all three cases, I will begin with a local source that prepares two qubits 1 and 2 in the state $|\psi\rangle{1,2}$ at time $t_1$, and then sends one qubit to Alice, and the other to Bob, who make randomly-selected space-like-separated measurements on their respective qubits at times $t_2$ and $t_3$, respectively.  Note that because these two measurements are space-like-separated, the temporal order of $t_2$ and $t_3$ may be different in different reference frames, and what we are truly indexing are the interaction events.  Alice and Bob meet to share their results at $t_4$.

The 0th example applies to any state $|\psi\rangle^{1,2}$, and for the other three cases, the source prepares the state,
\begin{equation}
|\psi\rangle^{1,2} = \frac{4}{5} |0\rangle^1 |0\rangle^2 + \frac{3}{5}|1\rangle^1 |1\rangle^2, \label{ExPsi}
\end{equation}
The preferred basis for this state is determined by the coupling interaction at the source, and it need not be the $\{|0\rangle, |1\rangle\}$ basis in which the state is represented in Eq. \ref{ExPsi}.

\emph{Example 0}:
\begin{figure}[h]
\begin{center}
\subfloat[][Alice's Measurement at $t_2$]{
\begin{tabular}{|c|c|c|c|}
  \hline
  Proportion & Relative World & Memory & History \\
 \hline
  $\textrm{Tr}(\rho^1 |a_+\rangle^1\langle a_+|^1)$ & $|a_+\rangle^1_{t_2}|a_+\rangle^A_{t_2}$ &  $|\psi\rangle_{t_1}^{1,2}$ & $\rho^1_{t_1}$  \\
  $\textrm{Tr}(\rho^1 |a_-\rangle^1\langle a_-|^1)$ & $|a_-\rangle^1_{t_2}|a_-\rangle^A_{t_2}$ & $|\psi\rangle_{t_1}^{1,2}$ & $\rho^1_{t_1}$\\
  \hline
\end{tabular}\label{T01}
}
\qquad
\subfloat[][Bob's Measurement at $t_3$]{
\begin{tabular}{|c|c|c|c|}
  \hline
Proportion & Relative World & Memory & History \\
 \hline
  $\textrm{Tr}(\rho^2 |b_+\rangle^2\langle b_+|^2)$ & $|b_+\rangle^2_{t_3}|b_+\rangle^B_{t_3}$ & $|\psi\rangle_{t_1}^{1,2}$& $\rho^2_{t_1}$ \\
  $\textrm{Tr}(\rho^2 |b_-\rangle^2\langle b_-|^2)$ & $|b_-\rangle^2_{t_3}|b_-\rangle^B_{t_3}$ & $|\psi\rangle_{t_1}^{1,2}$& $\rho^2_{t_1}$ \\
  \hline
\end{tabular}\label{T02}}
\qquad
\subfloat[][Alice and Bob Meet at $t_4$]{
\begin{tabular}{|c|c|c|c|}
  \hline
Proportion & Relative World & Memory & History \\
 \hline
  $\Big|\langle \psi |^{1,2}|a_+\rangle^1 |b_+\rangle^2\Big|^2$ & $|a_+\rangle^A_{t_4}|b_+\rangle^B_{t_4}$ & $|\psi\rangle_{t_1}^{1,2}$& $|a_+\rangle^1_{t_2}|a_+\rangle^A_{t_2}$, $|b_+\rangle^2_{t_3}|b_+\rangle^B_{t_3}$ \\
  $\Big|\langle \psi |^{1,2}|a_+\rangle^1  |b_-\rangle^2\Big|^2$ & $|a_+\rangle^A_{t_4}|b_-\rangle^B_{t_4}$ & $|\psi\rangle_{t_1}^{1,2}$& $|a_+\rangle^1_{t_2}|a_+\rangle^A_{t_2}$, $|b_-\rangle^1_{t_3}|b_-\rangle^B_{t_3}$ \\

  $\Big|\langle \psi |^{1,2}|a_-\rangle^1  |b_+\rangle^2\Big|^2$ & $|a_-\rangle^A_{t_4}|b_+\rangle^B_{t_4}$ & $|\psi\rangle_{t_1}^{1,2}$& $|a_-\rangle^1_{t_2}|a_-\rangle^A_{t_2}$, $|b_+\rangle^1_{t_3}|b_+\rangle^B_{t_3}$ \\

  $\Big|\langle \psi |^{1,2}|a_-\rangle^1  |b_-\rangle^2\Big|^2$ & $|a_-\rangle^A_{t_4}|b_-\rangle^B_{t_4}$ & $|\psi\rangle_{t_1}^{1,2}$& $|a_-\rangle^1_{t_2}|a_-\rangle^A_{t_2}$, $|b_-\rangle^1_{t_3}|b_-\rangle^B_{t_3}$ \\
  \hline
\end{tabular}\label{T03}}
\caption{These interaction tables show the proportion of lives with each history that experience each meeting event, which together define the different relative worlds that the lives of the two systems experience.}
\end{center}
\end{figure}

In order to highlight how the PL mechanism obeys local causality, I will first analyze a simplified example, without specifying a preferred basis.

First off, the physical density of lives that propagate through space-time from the source to Alice and Bob are given by the reduced density matrices of the two qubits, $\rho^1 =\textrm{Tr}_2 \big( |\psi\rangle^{1,2}\langle \psi|^{1,2}\big)$ and $\rho^2 = \textrm{Tr}_1 \big(|\psi\rangle^{1,2}\langle \psi|^{1,2}\big)$, and this naturally also determines how the lives of each qubit will separate in response to a local measurement.  Second, each life of each qubit carries a hidden note containing ``$|\psi\rangle^{1,2}$ '' --- the entangled state that was prepared locally at the source --- in its internal memory.

When Alice makes her measurement in the randomly-selected basis $\{ |a_+\rangle,|a_-\rangle \}$, her lives, and the lives of qubit 1, each separate into two relative worlds, with proportions $P(a_\pm) = \textrm{Tr}(\rho^1 |a_\pm\rangle\langle a_\pm|)$, as shown in Fig. \ref{T01}.  Likewise, when Bob makes his measurement, the proportions are $P(b_\pm) = \textrm{Tr}(\rho^2 |b_\pm\rangle\langle b_\pm|)$, as in Fig. \ref{T02}.  From this, it is clear that there is no correlation between the lives of Alice and Bob immediately after they make their measurements, since the joint probability distribution $P(a_\pm,b_\pm)$ is separable.

However, the internal memory state ``$|\psi\rangle_{t_1}^{1,2}$'' is passed from each life of qubit 1 to a life of Alice during her measurement, and from each life of qubit 2 to a life of Bob during his.  Finally, when Alice and Bob meet, the hidden notes locally govern the proportions with which lives of Alice meet lives of Bob, according to $P(a_\pm,b_\pm) = |\langle \psi|_{t_1}^{1,2} (|a_\pm\rangle^1 \otimes |b_\pm\rangle^2)|^2$ (see Fig. \ref{T03}), and thus the correct entanglement correlations are obeyed, while using only local information that propagates causally through space-time.

Now that the local causality of the mechanism has been explained, I will add the preferred basis to the discussion, making it clear exactly what relative world each life of each system is experiencing at all times.

\emph{Example 1}:

We begin with the prepared state of Eq. \ref{ExPsi}.  Suppose the preferred basis for both of these qubits is $\{|0\rangle, |1\rangle\}$.  This means that 16/25 of the lives of qubit 1 experienced a past meeting event with a life of qubit 2 in which both were in state $|0\rangle$, and 9/25 of the lives of qubit 1 experienced a past meeting event with a life of qubit 2 in which both were in state $|1\rangle$.  Likewise for the lives of qubit 2, which shared the experience of those same past events at $t_1$ on their respective world-lines.  All of these lives carry ``$|\psi_{t_1}\rangle$'' in their internal memories.

Now, consider an instance where both Alice and Bob happen to measure in the $\{|0\rangle, |1\rangle\}$ basis.  When Alice measures at $t_2$, 16/25 of her lives meet the 16/25 of the lives of qubit 1 in state $|0\rangle$, and 9/25 of her lives meet the 9/25 of the lives of qubit 1 in state $|1\rangle$ (Fig. \ref{T1}), and likewise when Bob measures qubit 2  at $t_3$ (Fig. \ref{T2}).  The internal memories of the lives of both the qubit and the measurer update to ``$|\psi_{t_2}\rangle = U^{1,A} |\psi_{t_1}\rangle^{1,2} |R\rangle^A$'' .  In this case, the lives of the qubit experience no change, since they are being measured in their already-preferred basis.  The situation for Bob and qubit 2 is symmetrically identical, with the updated internal memory state ``$|\psi_{t_3}\rangle = U^{2,B} |\psi_{t_1}\rangle^{1,2} |R\rangle^B$''.

When Alice and Bob meet to compare their results at $t_4$, the 16/25 of the lives of Alice with a $|0\rangle$ outcome meet the 16/25 of the lives of Bob with a $|0\rangle$ outcome, and likewise for the 9/25 of their lives with state $|1\rangle$, such that for all lives of Alice and Bob, the entanglement correlations of $|\psi\rangle$ have been obeyed (Fig. \ref{T3}).  This is mediated by the synchronized internal memory state ``$|\psi_{t_4}\rangle = U^{1,A} U^{2,B}|\psi_{t_1}\rangle^{1,2} |R\rangle^A|R\rangle^B$''.

\begin{figure}[h]
\begin{center}
\subfloat[][Alice's Measurement at $t_2$]{
\begin{tabular}{|c|c|c|}
  \hline
  Proportion & Experience & History \\
 \hline
  $16/25$ & $|0^1_{t_2}\rangle|0^A_{t_2}\rangle$ & $|0^1_{t_1}\rangle|0^2_{t_1}\rangle$, ``$|\psi_{t_2}\rangle$'' \\
  $9/25$ & $|1^1_{t_2}\rangle|1^A_{t_2}\rangle$ & $|1^1_{t_1}\rangle|1^2_{t_1}\rangle$, ``$|\psi_{t_2}\rangle$'' \\
  \hline
\end{tabular}\label{T1}
}
\qquad
\subfloat[][Bob's Measurement at $t_3$]{
\begin{tabular}{|c|c|c|}
  \hline
  Proportion & Experience & History \\
 \hline
  $16/25$ & $|0^2_{t_3}\rangle|0^B_{t_3}\rangle$ & $|0^2_{t_1}\rangle|0^1_{t_1}\rangle$, ``$|\psi_{t_3}\rangle$''  \\
  $9/25$ & $|1^2_{t_3}\rangle|1^B_{t_3}\rangle$ & $|1^2_{t_1}\rangle|1^1_{t_1}\rangle$, ``$|\psi_{t_3}\rangle$''   \\
  \hline
\end{tabular}\label{T2}}
\qquad
\subfloat[][Alice and Bob Meet at $t_4$]{
\begin{tabular}{|c|c|c|}
  \hline
  Proportion & Experience & History \\
 \hline
  $16/25$ & $|0^A_{t_4}\rangle|0^B_{t_4}\rangle$ & $|0^2_{t_3}\rangle|0^B_{t_3}\rangle$,$|0^1_{t_2}\rangle|0^A_{t_2}\rangle$,$|0^2_{t_1}\rangle|0^1_{t_1}\rangle$, ``$|\psi_{t_4}\rangle$'' \\
  $9/25$ & $|1^A_{t_4}\rangle|1^B_{t_4}\rangle$ &  $|1^2_{t_3}\rangle|1^B_{t_3}\rangle$,$|1^1_{t_2}\rangle|1^A_{t_2}\rangle$,$|1^2_{t_1}\rangle|1^1_{t_1}\rangle$, ``$|\psi_{t_4}\rangle$''  \\
  \hline
\end{tabular}\label{T3}}
\caption{These interaction tables show the proportion of lives with each history that experience each meeting event, which together define the different relative worlds that the lives of the two systems experience.  The history also contains the relative state in quotation marks, which has causally propagated in internal memory from past events to the present one.}
\end{center}
\end{figure}

\emph{Example 2}:
\begin{figure}[h]
\begin{center}
\subfloat[][Alice's Measurement at $t_2$ in the basis $\{|+\rangle,|-\rangle \}$]{
\begin{tabular}{|c|c|c|}
  \hline
  Proportion & Experience & History \\
 \hline
  $8/25$ & $|+^1_{t_2}\rangle|+^A_{t_2}\rangle$ & $|0^1_{t_1}\rangle|0^2_{t_1}\rangle$, ``$|\psi_{t_2}\rangle$''  \\
    $8/25$ & $|-^1_{t_2}\rangle|-^A_{t_2}\rangle$ & $|0^1_{t_1}\rangle|0^2_{t_1}\rangle$, ``$|\psi_{t_2}\rangle$''  \\
  $9/50$ & $|+^1_{t_2}\rangle|+^A_{t_2}\rangle$ & $|1^1_{t_1}\rangle|1^2_{t_1}\rangle$, ``$|\psi_{t_2}\rangle$'' \\
    $9/50$ & $|-^1_{t_2}\rangle|-^A_{t_2}\rangle$ & $|1^1_{t_1}\rangle|1^2_{t_1}\rangle$, ``$|\psi_{t_2}\rangle$''  \\
  \hline
\end{tabular}\label{T4}
}
\qquad
\subfloat[][Bob's Measurement at $t_3$ in the basis $\{|+\rangle,|-\rangle \}$]{
\begin{tabular}{|c|c|c|}
  \hline
  Proportion & Experience & History \\
 \hline
  $8/25$ & $|+^2_{t_3}\rangle|+^B_{t_3}\rangle$ & $|0^2_{t_1}\rangle|0^1_{t_1}\rangle$, ``$|\psi_{t_3}\rangle$''  \\
    $8/25$ & $|-^2_{t_3}\rangle|-^B_{t_3}\rangle$ & $|0^2_{t_1}\rangle|0^1_{t_1}\rangle$, ``$|\psi_{t_3}\rangle$''  \\
  $9/50$ & $|+^2_{t_3}\rangle|+^B_{t_3}\rangle$ & $|1^2_{t_1}\rangle|1^1_{t_1}\rangle$, ``$|\psi_{t_3}\rangle$''  \\
    $9/50$ & $|-^2_{t_3}\rangle|-^B_{t_3}\rangle$ & $|1^2_{t_1}\rangle|1^1_{t_1}\rangle$, ``$|\psi_{t_3}\rangle$''  \\
  \hline
\end{tabular}\label{T5}}
\qquad
\subfloat[][Alice and Bob Meet at $t_4$]{
\begin{tabular}{|c|c|c|}
  \hline
  Proportion & Experience & History \\
 \hline
    $784/2500$ & $|+^A_{t_4}\rangle|+^B_{t_4}\rangle$ & $|+^1_{t_2}\rangle|+^A_{t_2}\rangle$,$|+^2_{t_3}\rangle|+^B_{t_3}\rangle$,$|0^1_{t_1}\rangle|0^2_{t_1}\rangle$, ``$|\psi_{t_4}\rangle$''  \\
    $16/2500$ & $|+^A_{t_4}\rangle|-^B_{t_4}\rangle$ & $|+^1_{t_2}\rangle|+^A_{t_2}\rangle$,$|-^2_{t_3}\rangle|-^B_{t_3}\rangle$,$|0^1_{t_1}\rangle|0^2_{t_1}\rangle$, ``$|\psi_{t_4}\rangle$''  \\
    $16/2500$ & $|-^A_{t_4}\rangle|+^B_{t_4}\rangle$ & $|-^1_{t_2}\rangle|-^A_{t_2}\rangle$,$|+^2_{t_3}\rangle|+^B_{t_3}\rangle$,$|0^1_{t_1}\rangle|0^2_{t_1}\rangle$, ``$|\psi_{t_4}\rangle$''  \\
    $784/2500$ & $|-^A_{t_4}\rangle|-^B_{t_4}\rangle$ & $|-^1_{t_2}\rangle|-^A_{t_2}\rangle$,$|-^2_{t_3}\rangle|-^B_{t_3}\rangle$,$|0^1_{t_1}\rangle|0^2_{t_1}\rangle$, ``$|\psi_{t_4}\rangle$''  \\
    $441/2500$ & $|+^A_{t_4}\rangle|+^B_{t_4}\rangle$ & $|+^1_{t_2}\rangle|+^A_{t_2}\rangle$,$|+^2_{t_3}\rangle|+^B_{t_3}\rangle$,$|1^1_{t_1}\rangle|1^2_{t_1}\rangle$, ``$|\psi_{t_4}\rangle$''  \\
    $9/2500$ & $|+^A_{t_4}\rangle|-^B_{t_4}\rangle$ & $|+^1_{t_2}\rangle|+^A_{t_2}\rangle$,$|-^2_{t_3}\rangle|-^B_{t_3}\rangle$,$|1^1_{t_1}\rangle|1^2_{t_1}\rangle$, ``$|\psi_{t_4}\rangle$''  \\
    $9/2500$ & $|-^A_{t_4}\rangle|+^B_{t_4}\rangle$ & $|-^1_{t_2}\rangle|-^A_{t_2}\rangle$,$|+^2_{t_3}\rangle|+^B_{t_3}\rangle$,$|1^1_{t_1}\rangle|1^2_{t_1}\rangle$, ``$|\psi_{t_4}\rangle$''  \\
    $441/2500$ & $|-^A_{t_4}\rangle|-^B_{t_4}\rangle$ & $|-^1_{t_2}\rangle|-^A_{t_2}\rangle$,$|-^2_{t_3}\rangle|-^B_{t_3}\rangle$,$|1^1_{t_1}\rangle|1^2_{t_1}\rangle$, ``$|\psi_{t_4}\rangle$''  \\
  \hline
\end{tabular}\label{T6}}
\caption{These interaction tables show the proportion of lives with each history that experience each meeting event, which together define the different relative worlds that the lives of the two systems experience.    The history also contains the relative state in quotation marks, which has causally propagated in internal memory from past events to the present one.}
\end{center}
\end{figure}

Again, suppose the preferred basis for both of these qubits is $\{|0\rangle, |1\rangle\}$, but now consider an instance where Alice and Bob both measure the $\{|+\rangle = (|0\rangle +  |1\rangle)/\sqrt{2}, |-\rangle = (|0\rangle - |1\rangle)/\sqrt{2}\}$ basis.  Represented in this basis we have
 \begin{equation}
   |\psi\rangle = \frac{7}{10} |+^1\rangle |+^2\rangle +  \frac{1}{10} |+^1\rangle |-^2\rangle + \frac{1}{10}|-^1\rangle |+^2\rangle + \frac{7}{10}|-^1\rangle |-^2\rangle. \label{psiXX}
\end{equation}
Using the reduced density matrix we find that the half of the lives of Alice obtain a $|+\rangle$ and half obtain a $|-\rangle$ outcome --- although each of these outcomes has multiple histories, resulting in four different relative worlds (Fig \ref{T4}), and likewise for Bob (Fig \ref{T5}).  The evolution of the internal memories is of the same form as in Example 1.

For the 16/25 of the lives of qubit 1 that were in state $|0\rangle$, the measurement changes the preferred basis, and half of these lives change to $|+\rangle$ and the other half to $|-\rangle$.  Likewise for the 16/25 of the lives of qubit 2 in state $|0\rangle$.  For the 9/25 of the lives of qubit 1 in state $|1\rangle$, again half change to $|+\rangle$ and the other half to $|-\rangle$, and likewise for the corresponding lives of qubit 2.  In this case, the lives of the qubits experience Alice's and Bob's measurements as causing collapses, and they separate into four relative worlds.

When Alice and Bob meet to compare their results, 49/100 of Alice's lives with a $|+\rangle$ meet 49/100 of Bob's lives with a $|+\rangle$, 49/100 of their respective lives meet with $|-^A\rangle |-^B\rangle$, 1/100 with $|+^A\rangle |-^B\rangle$, and 1/100 with $|-^A\rangle |+^B\rangle$, such that the proper entanglement correlations of $|\psi\rangle$ have been obeyed.  Each of these four outcomes has multiple histories, resulting in the eight relative worlds shown in Fig. \ref{T6}.

For each history $h_i$ with total proportion $P_i$, there may be many specific future outcomes $f_{ij}$ for that history.  The proportion of lives from each history $h_i$ that experience each future $f_{ij}$ is given by,
\begin{equation}
P_{ij} = P_i |c_j|^2/(\sum_{k \in h_i}|c_k|^2),\label{Phistory}
\end{equation}
where the $c_j$ are the coefficients of each term (relative world) in the new preferred basis.  To determine which futures $f_{ij}$ belong to each history $h_i$, we consider how the measurement unitary $U^{1,A}$ acts on the product state representing each history.  For example, for the history $h_i = |0^1\rangle|0^2\rangle|R^A\rangle$, where $|R^A\rangle$ is Alice's `ready' state prior to the measurement, we have,
\begin{equation}
U^{1,A}|0^1\rangle|0^2\rangle|R^A\rangle = U^{1,A}\big(|+^1\rangle + |-^1\rangle\big)|0^2\rangle|R^A\rangle/\sqrt{2} = \big(|+^1\rangle|+^A\rangle + |-^1\rangle|-^A\rangle\big)|0^2\rangle/\sqrt{2}.
\end{equation}
From this we learn that both $|+^1\rangle|+^A\rangle$ and $|-^1\rangle|-^A\rangle$ are futures $f_{ij}$ of this history $h_i$  --- and their proportions $P_{ij}$ are given by the formula above, using coefficients $c_j$ from the state,
\begin{equation}
|\psi_{t_2}\rangle = U^{1,A}|\psi_{t_1}\rangle|R^A\rangle = \frac{4}{5\sqrt{2}}\big(|+^1\rangle|+^A\rangle +|-^1\rangle|-^A\rangle  \big)|0^2\rangle    +  \frac{3}{5\sqrt{2}}\big( |+^1\rangle|+^A\rangle -|-^1\rangle|-^A\rangle  \big)|1^2\rangle,
\end{equation}
recalling that ``$|\psi_{t_1}\rangle$'' is contained in the internal memories of all lives of qubit 1.

When Alice and Bob meet, neither of their preferred bases changes, and the lives with common history meet one another with proportions $P_{ij}$ determined by the coefficients $c_j$ in $|\psi\rangle$, represented in the new preferred basis, as in Eq. \ref{psiXX}.

\emph{Example 3}:
\begin{figure}[b]
\begin{center}
\subfloat[][Alice's Measurement at $t_2$]{
\begin{tabular}{|c|c|c|}
  \hline
  Proportion & Experience & History \\
 \hline
  $784/2500$ & $|0^1_{t_2}\rangle|0^A_{t_2}\rangle$ & $|+^1_{t_1}\rangle|+^2_{t_1}\rangle$, ``$|\psi_{t_2}\rangle$''  \\
    $441/2500$ & $|1^1_{t_2}\rangle|1^A_{t_2}\rangle$ & $|+^1_{t_1}\rangle|+^2_{t_1}\rangle$, ``$|\psi_{t_2}\rangle$''  \\
      $16/2500$ & $|0^1_{t_2}\rangle|0^A_{t_2}\rangle$ & $|+^1_{t_1}\rangle|-^2_{t_1}\rangle$, ``$|\psi_{t_2}\rangle$''  \\
    $9/2500$ & $|1^1_{t_2}\rangle|1^A_{t_2}\rangle$ & $|+^1_{t_1}\rangle|-^2_{t_1}\rangle$, ``$|\psi_{t_2}\rangle$''  \\
  $16/2500$ & $|0^1_{t_2}\rangle|0^A_{t_2}\rangle$ & $|-^1_{t_1}\rangle|+^2_{t_1}\rangle$, ``$|\psi_{t_2}\rangle$''  \\
    $9/2500$ & $|1^1_{t_2}\rangle|1^A_{t_2}\rangle$ & $|-^1_{t_1}\rangle|+^2_{t_1}\rangle$, ``$|\psi_{t_2}\rangle$''  \\
      $784/2500$ & $|0^1_{t_2}\rangle|0^A_{t_2}\rangle$ & $|-^1_{t_1}\rangle|-^2_{t_1}\rangle$, ``$|\psi_{t_2}\rangle$''  \\
    $441/2500$ & $|1^1_{t_2}\rangle|1^A_{t_2}\rangle$ & $|-^1_{t_1}\rangle|-^2_{t_1}\rangle$, ``$|\psi_{t_2}\rangle$''  \\
  \hline
\end{tabular}\label{T7}
}
\qquad
\subfloat[][Bob's Measurement at $t_3$]{
\begin{tabular}{|c|c|c|}
  \hline
  Proportion & Experience & History \\
 \hline
  $16513/62500$ & $|x^2_{t_3}\rangle|x^B_{t_3}\rangle$ & $|+^1_{t_1}\rangle|+^2_{t_1}\rangle$, ``$|\psi_{t_3}\rangle$''  \\
    $14112/62500$ & $|y^2_{t_3}\rangle|y^B_{t_3}\rangle$ & $|+^1_{t_1}\rangle|+^2_{t_1}\rangle$, ``$|\psi_{t_3}\rangle$''  \\
      $337/62500$ & $|x^2_{t_3}\rangle|x^B_{t_3}\rangle$ & $|+^1_{t_1}\rangle|-^2_{t_1}\rangle$, ``$|\psi_{t_3}\rangle$''  \\
    $288/62500$ & $|y^2_{t_3}\rangle|y^B_{t_3}\rangle$ & $|+^1_{t_1}\rangle|-^2_{t_1}\rangle$, ``$|\psi_{t_3}\rangle$''  \\
  $337/62500$ & $|x^2_{t_3}\rangle|x^B_{t_3}\rangle$ & $|-^1_{t_1}\rangle|+^2_{t_1}\rangle$, ``$|\psi_{t_3}\rangle$''  \\
    $288/62500$ & $|y^2_{t_3}\rangle|y^B_{t_3}\rangle$ & $|-^1_{t_1}\rangle|+^2_{t_1}\rangle$, ``$|\psi_{t_3}\rangle$''  \\
      $16513/62500$ & $|x^2_{t_3}\rangle|x^B_{t_3}\rangle$ & $|-^1_{t_1}\rangle|-^2_{t_1}\rangle$, ``$|\psi_{t_3}\rangle$''  \\
    $14112/62500$ & $|y^2_{t_3}\rangle|y^B_{t_3}\rangle$ & $|-^1_{t_1}\rangle|-^2_{t_1}\rangle$, ``$|\psi_{t_3}\rangle$''  \\
  \hline
\end{tabular}\label{T8}}
\qquad
\subfloat[][Alice and Bob Meet at $t_4$]{
\begin{tabular}{|c|c|c|}
  \hline
  Proportion & Experience & History \\
 \hline
    $7056/62500$ & $|0^A_{t_4}\rangle|x^B_{t_4}\rangle$ & $|0^1_{t_2}\rangle|0^A_{t_2}\rangle$,$|x^2_{t_3}\rangle|x^B_{t_3}\rangle$,$|+^1_{t_1}\rangle|+^2_{t_1}\rangle$, ``$|\psi_{t_4}\rangle$''  \\
    $12544/62500$ & $|0^A_{t_4}\rangle|y^B_{t_4}\rangle$ & $|0^1_{t_2}\rangle|0^A_{t_2}\rangle$,$|y^2_{t_3}\rangle|y^B_{t_3}\rangle$,$|+^1_{t_1}\rangle|+^2_{t_1}\rangle$, ``$|\psi_{t_4}\rangle$''  \\
    $7056/62500$ & $|1^A_{t_4}\rangle|x^B_{t_4}\rangle$ & $|1^1_{t_2}\rangle|1^A_{t_2}\rangle$,$|x^2_{t_3}\rangle|x^B_{t_3}\rangle$,$|+^1_{t_1}\rangle|+^2_{t_1}\rangle$, ``$|\psi_{t_4}\rangle$''  \\
    $3969/62500$ & $|1^A_{t_4}\rangle|y^B_{t_4}\rangle$ & $|1^1_{t_2}\rangle|1^A_{t_2}\rangle$,$|y^2_{t_3}\rangle|y^B_{t_3}\rangle$,$|+^1_{t_1}\rangle|+^2_{t_1}\rangle$, ``$|\psi_{t_4}\rangle$''  \\
    $144/62500$ & $|0^A_{t_4}\rangle|x^B_{t_4}\rangle$ & $|0^1_{t_2}\rangle|0^A_{t_2}\rangle$,$|x^2_{t_3}\rangle|x^B_{t_3}\rangle$,$|+^1_{t_1}\rangle|-^2_{t_1}\rangle$, ``$|\psi_{t_4}\rangle$''  \\
    $256/62500$ & $|0^A_{t_4}\rangle|y^B_{t_4}\rangle$ & $|0^1_{t_2}\rangle|0^A_{t_2}\rangle$,$|y^2_{t_3}\rangle|y^B_{t_3}\rangle$,$|+^1_{t_1}\rangle|-^2_{t_1}\rangle$, ``$|\psi_{t_4}\rangle$''  \\
    $144/62500$ & $|1^A_{t_4}\rangle|x^B_{t_4}\rangle$ & $|1^1_{t_2}\rangle|1^A_{t_2}\rangle$,$|x^2_{t_3}\rangle|x^B_{t_3}\rangle$,$|+^1_{t_1}\rangle|-^2_{t_1}\rangle$, ``$|\psi_{t_4}\rangle$''  \\
    $81/62500$ & $|1^A_{t_4}\rangle|y^B_{t_4}\rangle$ & $|1^1_{t_2}\rangle|1^A_{t_2}\rangle$,$|y^2_{t_3}\rangle|y^B_{t_3}\rangle$,$|+^1_{t_1}\rangle|-^2_{t_1}\rangle$, ``$|\psi_{t_4}\rangle$''  \\
    $144/62500$ & $|0^A_{t_4}\rangle|x^B_{t_4}\rangle$ & $|0^1_{t_2}\rangle|0^A_{t_2}\rangle$,$|x^2_{t_3}\rangle|x^B_{t_3}\rangle$,$|-^1_{t_1}\rangle|+^2_{t_1}\rangle$, ``$|\psi_{t_4}\rangle$''  \\
    $256/62500$ & $|0^A_{t_4}\rangle|y^B_{t_4}\rangle$ & $|0^1_{t_2}\rangle|0^A_{t_2}\rangle$,$|y^2_{t_3}\rangle|y^B_{t_3}\rangle$,$|-^1_{t_1}\rangle|+^2_{t_1}\rangle$, ``$|\psi_{t_4}\rangle$''  \\
    $144/62500$ & $|1^A_{t_4}\rangle|x^B_{t_4}\rangle$ & $|1^1_{t_2}\rangle|1^A_{t_2}\rangle$,$|x^2_{t_3}\rangle|x^B_{t_3}\rangle$,$|-^1_{t_1}\rangle|+^2_{t_1}\rangle$, ``$|\psi_{t_4}\rangle$''  \\
    $81/62500$ & $|1^A_{t_4}\rangle|y^B_{t_4}\rangle$ & $|1^1_{t_2}\rangle|1^A_{t_2}\rangle$,$|y^2_{t_3}\rangle|y^B_{t_3}\rangle$,$|-^1_{t_1}\rangle|+^2_{t_1}\rangle$, ``$|\psi_{t_4}\rangle$''  \\
    $7056/62500$ & $|0^A_{t_4}\rangle|x^B_{t_4}\rangle$ & $|0^1_{t_2}\rangle|0^A_{t_2}\rangle$,$|x^2_{t_3}\rangle|x^B_{t_3}\rangle$,$|-^1_{t_1}\rangle|-^2_{t_1}\rangle$, ``$|\psi_{t_4}\rangle$''  \\
    $12544/62500$ & $|0^A_{t_4}\rangle|y^B_{t_4}\rangle$ & $|0^1_{t_2}\rangle|0^A_{t_2}\rangle$,$|y^2_{t_3}\rangle|y^B_{t_3}\rangle$,$|-^1_{t_1}\rangle|-^2_{t_1}\rangle$, ``$|\psi_{t_4}\rangle$''  \\
    $7056/62500$ & $|1^A_{t_4}\rangle|x^B_{t_4}\rangle$ & $|1^1_{t_2}\rangle|1^A_{t_2}\rangle$,$|x^2_{t_3}\rangle|x^B_{t_3}\rangle$,$|-^1_{t_1}\rangle|-^2_{t_1}\rangle$, ``$|\psi_{t_4}\rangle$''  \\
    $3969/62500$ & $|1^A_{t_4}\rangle|y^B_{t_4}\rangle$ & $|1^1_{t_2}\rangle|1^A_{t_2}\rangle$,$|y^2_{t_3}\rangle|y^B_{t_3}\rangle$,$|-^1_{t_1}\rangle|-^2_{t_1}\rangle$, ``$|\psi_{t_4}\rangle$''  \\
  \hline
\end{tabular}\label{T9}}
\caption{These interaction tables show the proportion of lives with each history that experience each meeting event, which together define the different relative worlds that the lives of the two systems experience.  The history also contains the relative state in quotation marks, which has causally propagated in internal memory from past events to the present one.}
\end{center}
\end{figure}

Next we consider the case where $\{|+\rangle = (|0\rangle +  |1\rangle)/\sqrt{2}, |-\rangle = (|0\rangle - |1\rangle)/\sqrt{2}\}$ is the preferred basis, which means that 49/100 of the lives of qubit 1 experienced a past meeting event with a life of qubit 2 in which both were in state $|+\rangle$, 49/100 of the lives of qubit 1 experienced a past meeting event with a life of qubit 2 in which both were in state $|-\rangle$, 1/100 of the lives of qubit 1 were in state $|+\rangle$ and met a life of qubit 2 in state $|-\rangle$, and 1/100 were in state $|-\rangle$ and met a $|+\rangle$.  Likewise for the lives of qubit 2.

Now consider an asymmetric instance where Alice measures the $\{|0\rangle, |1\rangle\}$ basis, and Bob measures the basis $\{|x\rangle = (3|0\rangle +  4|1\rangle)/5, |y\rangle = (4|0\rangle - 3|1\rangle)/5\}$.  Represented in this measurement basis, we have,
 \begin{equation}
 |\psi\rangle = \frac{12}{25} |0\rangle |x\rangle +  \frac{16}{25} |0\rangle |y\rangle + \frac{12}{25}|1\rangle |x\rangle - \frac{9}{25}|1\rangle |y\rangle.
\end{equation}
Each qubit has four different histories prior to these measurements, and because both of their preferred bases change, the lives of each qubit are divided into eight relative worlds, as shown in Fig. \ref{T7} and Fig. \ref{T8}, respectively.  When Alice and Bob meet, the entanglement correlations of $|\psi_{t_1}\rangle$ are obeyed as usual, resulting in the 16 different relative worlds shown in Fig. \ref{T9}.

\textbf{The Local Entanglement Mechanism and Classical Observers}

Next we consider the example of an experimenter in a classical laboratory environment with a qubit prepared in the state $|\psi\rangle = (4|0\rangle + 3|1\rangle)/5$, and a measurement device in the state $|R\rangle$.  The experimenter then performs a measurement on the qubit in the $\{|0\rangle, |1\rangle\}$ basis, which causes the state of the qubit and measurement device to evolve to the entangled state,
 \begin{equation}
|\psi\rangle = \frac{4}{5} |0^q\rangle |0^m\rangle + \frac{3}{5}|1^q\rangle |1^m\rangle,
 \end{equation}
 where the superscripts $q$ and $m$ denote the qubit and measuring device, respectively.  This state is represented in the measurement basis by definition, which becomes the preferred basis for both systems after the measurement, as usual.

 Presuming for simplicity that $\{|0\rangle, |1\rangle\}$ was the preferred basis of the qubit, the 16/25 of its lives in state $|0\rangle$ and 9/25 in state $|1\rangle$ experience no change during the measurement.  The lives of the measurement device separate into two relative worlds, with 16/25 meeting a life of the qubit in state $|0\rangle$ and 9/25 meeting a life in state $|1\rangle$.

 The measurement device amplifies this result into the classical environment of the laboratory, and thus the states $|0^m\rangle$ and $|1^m\rangle$ effectively represent different classical worlds.  The only difference between this state and the two-qubit examples above is that it is very difficult in practice to measure the classical environment in any basis other than $\{ |0^m\rangle, |1^m\rangle \}$ --- which is part of the motivation behind various experimental attempts to create macroscopic cat states that may begin to approximate classical objects.  If the laboratory and qubit could be kept isolated from the external environment, then in principle they too could be moved apart and used to violate a Bell inequality, just as in the two-qubit case.

 Note that if the experimenter measures the qubit again using the same basis, the same 16/25 of the lives that met the lives of the qubit in state $|0\rangle$ before, simply meet the same 16/25 lives of the qubit again (although each individual life need not meet the same life as before), and likewise for the 9/25 with $|1\rangle$.  For all lives of the experimenter, this is consistent with having caused a collapse with the first measurement, which is the standard procedure for initializing a state.

 Now suppose that the experimenter has arranged for a Hadamard operation $H$ to be performed on the qubit before the second measurement.  Then the 16/25 of the experimenter's lives that met a $|0\rangle$ at the first measurement, now encounter the state $H|0\rangle  = (|0\rangle + |1\rangle)/\sqrt{2}$ at the second measurement, and 8/25 meet a life of the qubit with a $|0\rangle$ and 8/25 meet a life with $|1\rangle$.  The 9/25 of the experimenter's lives that met a $|1\rangle$ at the first measurement now encounter the state $H|1\rangle = (|0\rangle - |1\rangle)/\sqrt{2}$, and 9/50 meet a life of the qubit with a $|0\rangle$ and 9/50 meet a life with $|1\rangle$.

 The local entanglement mechanism works exactly the same way in all of these situations.  Perhaps the most striking lesson of PL is that if a classical observer in an entangled state could be measured in a basis other than $\{ |0^m\rangle, |1^m\rangle \}$, then that observer would experience a modified Born rule.  Consider Example 3 above, and note that for an observer in the state $|+\rangle$, the probability to experience ending up in states $|0\rangle$ or $|1\rangle$ is not 50\% --- as it would be for an isolated system in the state $|+\rangle$.  In PL, the lives of coherent quantum systems that can interact and become entangled in different bases experience these modified probabilities as a matter of routine.  However, it is important to note that measuring a classical system in such a basis would necessarily erase its classical memory of its previous $\{ |0^m\rangle, |1^m\rangle \}$ state, which makes it fundamentally impossible to record this effect experimentally (see \cite{frauchiger2016single} for a discussion of related topics).\\

%%%%%%%%%%%%%%%%%%%%%%%%%%%%%%%%%%%%%%%%%%%%%

\textbf{Weak Measurements and the Delayed Choice Quantum Eraser}

Weak measurements \cite{aharonov1988result, dressel2015weak} and the delayed-choice quantum eraser \cite{kim2000delayed, scully1982quantum} are fundamentally related experiments, both based on the interference between two wave packets mediated by entanglement with another system.  Letting the other system be a qubit, they are both concerned with measurements on a state of the form, $|\Omega\rangle = (\psi_1(x) |0\rangle + \psi_2(x)|1\rangle)/\sqrt{2}$, where $\psi_1(x)$ and $\psi_2(x)$ are non-orthogonal mostly-overlapping Gaussian wave packets.

In the case of a weak measurement, $\psi_1(x)$ and $\psi_2(x)$ originate from the same broad pointer function, translated by a very small amount in opposite directions by the coupling Hamiltonian $H = g(t) \sigma_z \hat{P}$, and the interference between them will reveal the weak value $ (\sigma_z)_w = \langle \phi| \sigma_z |\psi\rangle / \langle \phi | \psi \rangle$ for preselection $|\psi\rangle = (|0\rangle + |1\rangle)/\sqrt{2}$ and a general post-selection $|\phi\rangle$ on the qubit (where $g(t)$ is the coupling strength, $\sigma_z = |0\rangle\langle 0| - |1\rangle\langle 1|$ is a qubit operator, and $\hat{P}$ is the momentum operator of the pointer).

For the delayed-choice quantum eraser, $\psi_1(x)$ and $\psi_2(x)$ are beam profiles centered at the same point on a common screen, with opposite phase tilt $e^{\pm i k x}$, and they reveal a maximally-visible interference pattern if the qubit is post-selected as $|\phi\rangle = (|0\rangle \pm |1\rangle)/\sqrt{2}$, and no visible interference pattern at all if it is post-selected as $|\phi\rangle = |0\rangle$ or $|\phi\rangle = |1\rangle$.  The apparent paradox of this experiment is that the data can be recorded on the screen long before the choice of measurement for the qubit is determined, and yet upon post-selection, each data point on the screen seemed to already correspond to an interference pattern, or the lack thereof, \emph{before} this choice was made.  A similarly paradoxical effect appears for the weak measurement.

The resolution of the paradox is straightforward in PL.  When the screen is measured, we use the reduced density matrix of the entangled state to determine how the lives of the continuous system become separated into different worlds at each location $x$, which results in a distribution $P(x) = (|\psi_1(x)|^2 + |\psi_2(x)|^2)/2$ of lives at location $x$ on the screen --- which contains no interference fringes at all.  All of these lives propagate carrying the hidden information of $|\Omega\rangle$.  Likewise, when the qubit is measured, we use its reduced density matrix, which causes the lives of the qubit to separate into equal proportions, regardless of which basis is measured.  In any event, the lives of the qubit also carry the hidden information $|\Omega\rangle$.  Finally lives of the devices which measured the qubit and the screen meet, and the entanglement correlations in $|\Omega\rangle$ now play their role, determining the proportion of lives at each location $x$ that will meet with lives in each relative world of the qubit.

To be specific, suppose the qubit was measured in the $\{|+\rangle = (|0\rangle + |1\rangle)/\sqrt{2}, |-\rangle = (|0\rangle - |1\rangle)/\sqrt{2} \}$ basis, and its lives divided evenly into two relative worlds.  Represented in this basis we have,
\begin{equation}
|\Omega\rangle = G(x) \cos(kx) |+\rangle + i G(x) \sin(kx) |-\rangle,
\end{equation}
where $P(x) = G(x)^2$ is a normalized Gaussian function, and thus when the two systems meet, the lives of the qubit in relative world $|+\rangle$ meet the distribution $G(x)^2 \cos^2(kx)$ of lives of the screen, and the lives of the qubit in relative world $|-\rangle$ meet the distribution $G(x)^2 \sin^2(kx)$.

Again, there was no interference pattern in $P(x)$ --- the interference patterns did not emerge until the lives of both systems met again and the entanglement correlations are obeyed --- and then only if the qubit was measured in particular bases.

If instead the qubit had been measured in the $\{|0\rangle, |1\rangle \}$ basis, and its lives were divided evenly into these two relative worlds, then there would be no interference patterns.  In this basis we have,
\begin{equation}
|\Omega\rangle = G(x) e^{ikx} |0\rangle + G(x) e^{-ikx} |1\rangle,
\end{equation}
and thus when the two systems meet, the lives of the qubit in the relative world $|0\rangle$ meet the distribution $G(x)^2/2$ of lives of the screen, and the lives of the qubit in relative world $|1\rangle$ meet a separate distribution $G(x)^2/2$, and thus no interference pattern emerges.\\

\textbf{Remote Entanglement}

The Loophole-Free Bell Test performed at Delft \cite{hensen2015loophole} creates entanglement between the energy eigenstates of two remotely located diamond NV-centers that do not directly interact, and the mechanism for this is worth elucidating in PL.  We will examine a somewhat idealized version of the experiment, which takes place in two steps, which are repeated many times.

First, the two diamond qubits are initialized in their first excited states.  Both qubits have identical probability to decay into their ground state and emit a single photon.  The photons are directed toward each other to a common beam splitter in the middle, and detected at either output port of the beam splitter, using detectors that are sensitive to photon number.  Second, the two qubits are measured in randomly selected contexts, chosen for an ideal Bell test.  This measurement is made before the emitted photons, travelling at $c$, would have time to arrive at the beam splitter in the middle.  The data from many runs are collected, and only runs where a single photon was detected are considered, which projects the state of the qubit onto a Bell state.  Not only do the two qubits never interact, but in a given run, even after the photon has been detected in the middle, there has still not been time for an influence traveling at $c$ to have reached the other qubit --- and the qubits are measured even before this anyway.  Nevertheless, the two qubits are entangled, and their measured correlations violate the Bell inequality.

Formally, the entangled state of each diamond qubit with the electromagnetic field is, $a|f\rangle|0\rangle + b|g\rangle|1\rangle$, where $|f\rangle$ and $|g\rangle$ are respectively, the first excited state and the ground state of the diamond qubit, $|1\rangle$ and $|0\rangle$ are respectively, the state of one photon and the vacuum state of the field, and $a$ and $b$ are normalized coefficients related to the probability of decay.  Taking the direct product of two such states and letting the field modes sum (at the beam splitter) gives $|\psi \rangle =  a^2|ff\rangle|0\rangle + b^2|gg\rangle|{2}\rangle + ab(|gf\rangle + |gf\rangle)|{1}\rangle$.  Projecting onto $|{1}\rangle$ (one photon) and renormalizing gives the Bell state $(|gf\rangle + |gf\rangle) / \sqrt{2}$ of the two diamond qubits.

In PL, we begin by taking the reduced density matrix of each qubit, $|a|^2 |f\rangle \langle f| + |b|^2 |g\rangle \langle g|$, which shows how the lives of each qubit separate when they are measured.  Likewise, the reduced density matrix of the field tells us how the lives divide at the beam splitter and detectors.  When the lives of these systems meet (the coincidence counting of the measuring devices for the qubits, and the photon detectors), the correlation rules hidden in the internal memories of the lives of each system ensure the entanglement correlations of the state $|\psi \rangle$ are obeyed.\\

\textbf{Relative Worlds and Local Interactions Between Lives}

The most elementary physical systems in PL are excitations of the quantum field, which always comprise a specific subset of the Nonstandard infinity of lives of the field.  Lives are never separated into finite groups, always belonging to a continuum (or sub-continuum) of lives of a system --- only the relative proportions between different sub-continua are finite rational fractions.

In general, lives of the field in the same relative world interact locally with one another when they meet at the same location, but for lives of the same system, these interactions never produce observable outcomes in the external memories of these lives --- which is to say that they never cause the lives to experience collapse as they separate into different relative worlds, and thus the individual lives are unaware of these interactions.  This produces the usual wave-like unitary evolution observed in isolated quantum systems, which again bears a striking resemblance to \emph{many-interacting-worlds}\cite{hall2014quantum} and hydrodynamical \cite{madelung1926hydrodynamical, trahan2005multidimensional, schiff2012communication} models.

When two different systems interact, their lives meet one-to-one at events in space-time, which implies that each system must have the same total number of lives for consistency --- limited by a particular maximum Nonstandard infinity for excitations.  In these meetings, the interaction produces outcomes that are recorded in the external memory of each life, which may cause the lives to separate into different relative worlds.  The discrete set of outcomes in the past interaction cone of a given life fully defines that life's experience of its relative world --- and each discrete outcome explicitly requires an interaction with another system.  Thus, the complete experience of a relative world is truly a discrete list of correlations with other systems.

To help clarify the one-to-one interaction rule, suppose that a neutron moves along a trajectory where it will be ballistically incident with certainty on another neutron which is at rest in a superposition of two different positions along the trajectory of the first neutron, with half the lives of the target neutron at location 1 and the other half at location 2.  When the lives of the ballistic neutron arrive at location 1, half of them encounter the lives of the target neutron that are there, leaving the other half unaffected.  When the remaining half arrive at location 2, they encounter the lives of the neutron there, completing the one-to-one matching.  Notice that for the lives of the ballistic neutron, it seems as though the target neutron has collapsed nonlocally into one of two locations, but really the lives were there all along.  The final result is an entangled state of the two neutrons, whose correlation rules are now carried in the internal memories of the lives of the two particles.

\begin{figure}
\begin{center}
\caption{(Color Online) A ballistic collision between two particles with no long-range interaction is depicted below.  Everything exists in the same space-time, and the extra copies of the $x$-axis serve only to show distinguish lives of the two particles with different memories.  (left) The two particles begin well-separated, moving toward each other in mirror symmetry.  (center) During the ballistic collision of the two particles, each life of one particle meets a life of the particle meet at an event in the center, and they both encode the new state ``$|\psi\rangle = \big(|B\rangle^1|B\rangle^2 + |T\rangle^1|T\rangle^2\big)/\sqrt{2}$'' into their internal memories.  Furthermore, half of them experience a bounce and encode $|B\rangle^1|B\rangle^2$ into their external memories, and the other half experience transmission, and encode $|T\rangle^1|T\rangle^2$ in their external memories. (right) The two particles after the collision has ended, now with all lives carrying the entangled state ``$|\psi\rangle$'' in their internal memories. }
\includegraphics[width=6.5in]{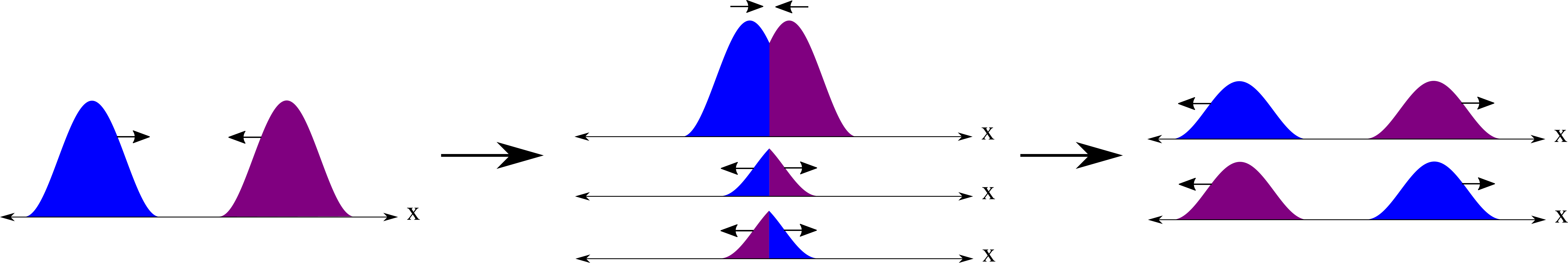}\label{BallScatter}
\end{center}
\end{figure}

In order to emphasize that the lives of different systems truly interact at events in space time, we consider an idealized case of symmetric ballistic scattering of two particles in 1D with no long-range interaction, in which each particle has a 50\% chance to bounce of the other, and a 50\% to transmit through the other, resulting in a maximally entangled state due to conservation of momentum.  Suppose the two packets begin completely separated, as shown in Fig. \ref{BallScatter}, and then at each instant of time, some lives of each particle move into the same location as lives of the other and meet at an event in space-time.   At these events, the internal memories of those lives synchronize to $\big(|B\rangle^1|B\rangle^2 + |T\rangle^1|T\rangle^2\big)/\sqrt{2}$, where $|B\rangle$ represents a bounce, and $|T\rangle$ represents a transmission.  Likewise, half of the lives of each particle experience bouncing ballistically off one another, encoding $|B\rangle^1|B\rangle^2$ in external memory, and the other half experience transmission, encoding $|T\rangle^1|T\rangle^2$ in external memory.
For this symmetric example, the number of lives of each particle arriving at the center per unit time is always the same, but if the situation is asymmetric, and the numbers of lives of each system in that location are not equal, then the excess lives do not interact at all, and nothing is encoded in their memories.  Note also that scattering with long-range coupling is more complicated because the intermediate field must also be treated as a system comprising lives which carry the causal influence from one particle to the other.

It may be that lives of different systems can also interact locally without producing observable outcomes, just like lives of the same system.  In either case, the property that allows lives of the same system to be distinguished from lives of different systems is just another correlation rule hidden in their internal memory.  Whenever an excitation is created from the field, all of the lives of the systems encode information in their internal memory which identifies them with that specific system, and which then mediates future local interactions between lives.  One speculative example could be that when an electron-positron pair is created from the vacuum field, a group of lives of the field acquire mass, charge, etc., and a message in internal memory that says something like ``I am a life of the electron (positron) that was created at event X in spacetime.''

This is consistent with the most general definition of a life within this axiomatic framework: A point-like object which carries information that governs how it interacts with other point-like objects at interaction events in space-time.  The carried information includes the correlation rules hidden in internal memory, which have been the central themes of this article, but it also includes properties like mass, charge, spin, kinetic energy, position, momentum, etc., which were already well-understood to govern interactions between physical systems.

At an even deeper level of abstraction, even the position of a life in 3-space could be recast as just another internal piece of information that the life carries, which governs when interaction events occur with other lives.  Without physical 3-space, all lives coexist at the single point in a 0-space, and an interaction event between two lives occurs when they carry the same values of position $\vec{x}$ at the same moment in time, and is otherwise governed by all of the other carried information as usual.\\

\textbf{The Aharonov-Bohm (AB) Effect}

Consider the AB effect \cite{aharonov1959significance} in a Mach-Zehnder interferometer with a solenoid enclosed by the arms, traversed by a charged particle.  The flux in the solenoid affects the relative phase of the interference between the two arms, even though the field is zero everywhere the particle travels.

The axioms of PL do not allow action-at-a-distance, and thus the simplest viable explanation of the AB effect is that lives of the field in the vacuum state propagate carrying phase information about the shape of the field in their internal memories, radiating outward from the solenoid at or below $c$, even in regions where the classical field strength is zero.  This hidden information is collected locally at interactions events between the lives of the field and the lives of the particle as they travel the arms of the interferometer, and governs the interference correlations at the second beam splitter.  The propagation of this hidden information in the electromagnetic field is a purely quantum feature that has no analog in classical electrodynamics --- in particular it need not correspond to the classical electric potential or vector potential.\\

\textbf{The status of $\Psi$ and $\psi$ in PL}

The barest interpretation says that quantum theory is nothing more or less than the best calculational tool available for an agent on a world-line to update their beliefs about the world when they receive new information, and the updates to prior belief are properly obtained using Bayes' theorem.  In PL, each individual life defines an experience by the list of past collapse events on its world-line, and each life carries its own relative wavefunction $\psi$ in its internal memory, which is updated as it propagates through space-time and interacts.  The relative wavefunction of a given life only makes predictions about events that may occur along that life's future world-line, but never about remote events.  As such, the ontology of PL is consistent with the epistemology of the Quantum Bayesian Interpretation (QBism) of quantum mechanics \cite{fuchs2011quantum, fuchs2013quantum, fuchs2014introduction}, at the level of the experience of individual lives.

Consider again the Bell test example with the entangled state $|\psi\rangle = (|00\rangle + |11\rangle) /\sqrt{2}$.  If Alice measures in the $\{ |0\rangle, |1\rangle \}$ basis, her lives divide equally into a world where her updated relative wavefunction is $|00\rangle$ and another where it is $|11\rangle$, which corresponds to an immediate local measurement outcome for Alice; but Alice does not learn anything at all about Bob's remote state.  Instead, the relative wavefunction $|00\rangle$ predicts that if she receives a signal from Bob in the future, that signal will be consistent with Bob having had a state $|0\rangle$ --- and it is only at that future meeting event that entanglement correlations will be obeyed for Alice.  For Bob, the story is the same, and thus the lives of Bob that get $|0\rangle$ predict they will later learn that Alice also got a $|0\rangle$, and vice versa, and when the two do exchange signals in the future, both of their predictions are satisfied --- the lives of Alice with $|0\rangle$ meet the lives of Bob with $|0\rangle$, and likewise for their lives with $|1\rangle$.

Each life carries a relative wavefunction $\psi$ in its internal memory, and collectively, all $\psi$ of all lives in the universe define an objective universal wavefunction $\Psi$ --- in a particular reference frame --- which undergoes unitary evolution to nonlocally track the correlations during all local interactions in the universe.  $\Psi$ is only an objective bookkeeping tool, which no individual life ever has access to, and rather it is the different $\psi$ in the memories of lives which belong to the fundamental ontology of PL.  And again, the $\psi$s hidden in the local memories of the lives throughout space-time also have the added benefit of providing a Lorentz-invariant narrative, which the universal wavefunction $\Psi$ does not.

In the language of Spekkens and Harrigan \cite{harrigan2010einstein}, PL is \emph{$\Psi$-ontic}, since in a given frame, the pure state of the universal wavefunction $\Psi$ is uniquely determined by the configuration of the lives throughout space-time, including their internal memories.  This classification may be somewhat dubious, since it refers to hidden variable models which give the probability distribution to obtain a single definite outcome for a given measurement, rather than predicting that all of those outcomes occur for different proportions of lives --- which is the true ontic state in PL.  Nevertheless, the set of predicted proportionalities for the outcomes of all possible measurements uniquely identifies each distinct $\Psi$, which still seems to fall under the umbrella of \emph{$\Psi$-ontic}.

Furthermore, PL is neither \emph{$\Psi$-complete} nor \emph{$\psi$-complete}, since neither $\Psi$ nor $\psi$ specifies the preferred basis, and $\Psi$ is also
agnostic to the dynamics of individual lives in space-time, both of which are essential for the definition of continuous world-lines ($\psi$ encodes at least a phase based on the trajectory of the life).  Specifically, even given the wavefunction $\Psi$ and the preferred basis, taking the reduced density matrices of every individual system reveals the collective distribution of all lives in space-time, but this still does not reveal the trajectories that the individual lives follow.

If the dynamics for individual lives is deterministic, then it seems plausible that two lives in the continuum, with different, but non-orthogonal internal wavefunctions, could have the exact same mix of noncontextual and contextual hidden variables predicting which outcome they will experience for all possible measurements.  This would make the experiences of the individual lives \emph{$\psi$-epistemic} in the true sense of Spekkens and Harrigan.\\

\textbf{Speculations on Gravity}

In order to claim that this is a complete ontological framework for nature, we must also consider how it might handle a theory of gravity.  The fact that PL is an ontology of fundamental point-like objects in space-time --- which obey all of the constraints of special relativity and local causality, instead of a theory of a fundamental universal wavefunction in configuration space --- could open the door to a new type of unification between general relativity and quantum mechanics.

One deep conflict between quantum mechanics and general relativity is that it seems impossible to consistently define a wavefunction on a superposition of space-times with different topologies, but the (as yet unknown) local dynamical equations that govern the motion of individual lives should still be applicable, and could lead to a formal treatment for massive superpositions.

One further simplification, which has already been explored extensively \cite{doran1993gravity, lasenby1993cosmological, lasenby1998gravity, hestenes2005gauge, hsu2006yang, hestenes2008gauge, hsu2011unified, jong2012model, hsu2012space}, is to assume that space-time has a trivial topology.  Any curved space-time with a trivial topology can be continuously geometrically deformed into a flat shape, which means that the physical effects of curved space-time predicted by general relativity can be replaced by an effective potential on a flat space-time with a Minkowski metric.  Then a massive superposition state can be treated as a superposition of different effective potentials embedded in the same flat space-time instead of as a superposition of differently-curved space-times.

Using this picture, the quantum gravity field could be treated no differently than the other quantum fields in PL, and the effective potentials would reproduce all of the observed effects of general relativity.  For a given effective potential, the lives of the quantum gravity field would propagate through the objective flat space-time carrying all of the effects of the gravitational potential.  For a massive superposition, the lives of the field would propagate carrying different effective potentials, and when the superposed gravity field interacts with another system, they could become entangled, and the lives of the field and the system would divide into correlated relative worlds just as in any other interaction in PL.\\

%%%%%%%%%%%%%%%%%%%%%%%%%%%%%%%%%%%%

\textbf{Experience and Free Will}

Each life of a system follows a particular world-line which defines the relative world experienced by that life, and the continuum of lives therefore defines a continuum of different perspectival experiences of the same universe.  Each life encodes a memory of the outcomes and correlations of a discrete set of interaction events within its past interaction cone, and it is this classical memory which defines the relative world experienced by that life.

In the classical regime of many densely interacting systems, the lives of each system experience relative worlds that are so similar that they are well-approximated by a single common world experience.  This explains the classical experience of macroscopic systems embedded in classical environments.

While individual lives record a memory of their experience, individual lives are not conscious observers, and conscious observers play no special role in PL.  Human consciousness seems to emerge from the frequent and repeated interactions of many physical systems within the brain, and thus it appears to fall into the classical realm.  If the brain can somehow maintain and manipulate coherent quantum states in isolation from this densely-interacting environment, the specific mechanism remains unknown.

We do not yet know the if correct dynamics for all individual lives is deterministic.  The collective distribution of all lives is constrained by unitary evolution to be deterministic, but random events remain plausible in the individual dynamics of lives, so long as the collective distribution is obeyed.

Either way, the conscious experience of free choice is arguably indistinguishable from the experience of weighted randomness, with the freedom originating in consciously choosing the weights of different outcomes prior to random selection of just one.  In this sense, a conscious observer experiences free choice whenever an observable interaction occurs and its systems' lives separate into different relative worlds --- i.e. whenever it experiences a collapse event.

If we entertain the idea that the brain does somehow maintain and manipulate isolated quantum systems, then if the dynamics of lives is deterministic, the free will of experiencing collapse events is simply an illusion created by the inability of lives to self-locate within their continuum (i.e. they cannot foresee which outcomes they will experience, even though it is deterministic).  On the other hand, if the dynamics is nondeterministic, then this randomness could be the most fundamental source of free will within the brain.

A final possibly remaining source of free will would allow that lives themselves can somehow choose which relative worlds they will `collapse' into --- which is sometimes called \emph{agency}.  More precisely, if a life that is headed to relative world $|0\rangle$ could somehow force a swap with another life headed to relative world $|1\rangle$ (such that the overall marginal distribution is unchanged), then this would be literal free will at the level of individual lives.  It is important to note that the ability to control which random outcome one will experience in a quantum measurement would enable violation of the No-Signalling bound \cite{cirel1980quantum, popescu1994quantum} using the PL entanglement mechanism --- although this would be the same for almost any interpretation of quantum mechanics.  I have assumed throughout this article that lives lack agency, since there is no physical evidence to suggest that this type of freedom exists in nature.  Nevertheless, it remains a philosophical possibility.\\

\textbf{Summary}

While incomplete, the PL model represents a new level of unification between special relativity and quantum mechanics, and makes new physical predictions about the speed of the dynamical collapse process.

All of the nonclassicality of quantum physics emerges at the level of individual lives and the relative worlds they experience --- specifically the experiences of random collapses with Born rule statistics, contextuality, and entanglement correlations.  The classical regime only emerges in densely-interacting environments due to the local entanglement-correlation mechanism of PL.  Furthermore, the different relative worlds that lives split into during an interaction are defined by the preferred eigenbasis of the measurement, making PL explicitly context-dependent, but while the proportion of lives that experience each outcome is noncontextual, the experience of the individual lives is not --- in agreement with the Bell-Kochen-Specker theorem \cite{bell1966problem, kochen1967problem}.  And most importantly, this model explains how the lives of systems involved in repeated interactions can experience the violation of Bell inequalities without any nonlocal, atemporal, or otherwise bizarre, causal influences.

This is a work in progress, and I am still considering many alternative models which are variants of this one.  Hopefully I have hinted at some of the directions that these alternatives might follow.  I want to emphasize this because I am not yet strongly committed to many of the details presented in this article --- I will be happy with any locally causal model of quantum mechanics in space-time.  The largest and most important remaining challenge is to find the proper relativistic dynamical equations for the lives of the universe (or even for a simple system), and to establish whether or not they are deterministic --- and finding candidate dynamical equations should also help to narrow down the possible variations.  With great luck, imposing relativistic constraints on the individual lives will also lead to the continuous solutions that are currently missing from quantum field theory --- possibly by including the effects of a quantum gravity field.

I hope that I have introduced the PL framework coherently and comprehensibly, and that this article will pique other researchers' interest in this physical model and the many unsolved problems it presents.

Given that every life in PL is a quasi-classical point-like object with definite physical properties in a preferred basis, which propagates through the universal space-time at or below $c$ and interacts only at local events along its world-line, PL tells an easily visualized quasi-classical objective story about nature, of the sort I imagine Einstein would have liked.  I also believe that this makes it an extremely useful tool for teaching quantum mechanics and making sense of some of its apparently paradoxical aspects --- for example through the classroom exercise presented in the Appendix.\\

\textbf{Acknowledgments:}  I would like to thank all of the following researchers for humoring me through many discussions as these ideas solidified.  In no particular order, they are:  Walter Lawrence, David Cyganski, Justin Dressel, Matt Leifer, Kevin Vanslette, Luis Pedro Garc\'{\i}a-Pintos, Kelvin McQueen, Roman Buniy, Paul Raymond-Robichaud, Yakir Aharonov, Jeff Tollaksen, Taylor Lee Patti, Travis Norsen, and Gregg Jaeger.   This research was supported (in part) by the Fetzer Franklin Fund of the John E. Fetzer Memorial Trust.

\bibliographystyle{ieeetr}
\bibliography{Parallel_Lives_Bib}

\section{Appendix}

There are many details that one must become comfortable with before the elegance of this model becomes obvious, and toward this end, I have devised a classroom exercise for a group of students that will allow them to participate in a simulation of the local entanglement mechanism themselves, and violate a Bell inequality.

Some of the students in the exercise will acts as lives of different systems, while others will play the role of nature, and act as referees, who read and write from the memories of different students, and determine which lives can meet.  In general, we will need the same number of students to acts as lives for every physical system we want to simulate, and we can keep the minimum total number small by restricting ourselves to only a few simple measurement settings --- since each student technically represents an infinite number of lives, and we need to divide the lives into the right proportions.  To simulate space-like separation, students cannot communicate directly during the exercise.

Each student who represents lives of a system will carry a notebook.  The notebook contains a label for which system they belong to, like `qubit 1,' and they also contain pages for internal memory, and other pages for external memory.  Due to the initial interaction of the two systems, each student starts off with a pure quantum state $|\psi\rangle$ in their internal memory, which includes every system in the past interaction cone of the system.  Furthermore, each student has one of the outcomes $|a\rangle^1|b\rangle^2$ of the interaction encoded in their external memory (which defines the preferred basis), along with a history of outcomes of past interactions for all other systems $|c\rangle^o$ in the past interaction cone.  The proportion of students in each of these relative worlds is $P(a,b,c) = |\langle \psi |a\rangle |b\rangle |c\rangle |^2$.

Now, suppose that system 1 interacts via unitary $U$ with a new system 3, and the students of each system meet one-to-one.  The referee now reads the internal memories $|\psi_1\rangle$ and $|\psi_3\rangle$ of an interacting pair of students from each system, and then updates them both to contain the relative state $|\psi'\rangle = U |\psi_1\rangle |\psi_3\rangle$.  The interaction may also cause the preferred basis to change, and then each student encodes an outcome $|x\rangle^1|y\rangle^3$ in that basis into their external memory (which determines which relative world they experience), and the pair shake hands to signify their meeting event.  Just as before, the proportion of students in each relative world is given by $P(x,y,z) = |\langle \psi |x\rangle |y\rangle |z\rangle |^2$, where again $|z\rangle^o$ is the history state of other systems, which includes the outcome of the previous interaction between systems 1 and 2, and the referee pairs the students off according to these proportions.

When macro-scale systems meet, the interaction unitary is identity, and the preferred bases do not change, but their lives still synchronize their internal memories and encode the meeting into their external memories.  These are the only rules we need in order to simulate a Bell experiment.

Let us consider one of the classic examples, developed by Wigner \cite{wigner1970hidden} and Mermin \cite{mermin1985moon}, of a Bell test involving a source which prepares two qubits in the singlet state $|\psi_0\rangle^{1,2} = \big(|0\rangle^1|1\rangle^2 - |1\rangle^1|0\rangle^2  \big)/\sqrt(2)$, and sends one to Alice and one to Bob.  We assume that we can ignore other systems in the interaction history of these two at the beginning of this experiment.  At space-like separation, Alice and Bob randomly choose among the three equally spaced angles in the $XZ$-plane and measure the qubit along that axis.  For this simulation we will need 16 students to act as lives, and 3 to acts as local referees.  We will begin with this state already prepared, and let $\{|0\rangle,|1\rangle \}$ be the preferred basis for both systems.  Each qubit will have 8 students representing it, and this means that 4 students of qubit 1 are in state $|0\rangle^1$ have a record in their external memory of meeting a student of qubit 2 in state $|1\rangle^2$, and the other four are in state $|1\rangle^1$ and met a student of qubit 2 in state $|0\rangle^2$.  All 16 of them have $|\psi_0\rangle^{1,2}$ written in internal memory.

Now, the 8 students of qubit 1 walk over to Alice, and the 8 students of qubit 2 to Bob.  For simplicity, we let setting 1 be the already-preferred basis, and settings 2 and 3 be, $\{\frac{\sqrt{3}}{2}|0\rangle + \frac{1}{2}|1\rangle, \frac{1}{2}|0\rangle - \frac{\sqrt{3}}{2} |1\rangle \}$, and $\{\frac{\sqrt{3}}{2}|0\rangle - \frac{1}{2}|1\rangle, \frac{1}{2}|0\rangle + \frac{\sqrt{3}}{2} |1\rangle \}$, respectively.  We will let Alice begin in the ready state $|R\rangle^A$, and the interaction unitary is then $U^{1,A} = |s_1\rangle^A |s_1\rangle^1 \langle R|^A \langle s_1|^1 + |s_2\rangle^A |s_2\rangle^1 \langle R|^A \langle s_2|^1 +$ (terms with $\langle R_\bot |^A$), where $|s_1\rangle$ and $|s_2\rangle$ are the basis states of the measurement setting, and similar for Bob's $U^{2,B}$.

If Alice chooses setting 1, then 4 of her students meet a student of qubit 1 already in state $|0\rangle^1$ and experience outcome $|0\rangle^A$, and the other 4 meet a student already in state $|1\rangle^1$ and experience outcome $|1\rangle^A$.  Using $P(s_1^1, s_1^A, h^2)$, we see that if Alice measures either setting 2 or 3, then three students of qubit 1 which were previously in the relative world $|0\rangle^1$ now enter relative world $|s_2\rangle^1$ and meet students of Alice in state $|s_2\rangle^A$, one student of qubit 1 which was previously in the relative world $|0\rangle^1$ now enters relative world $|s_1\rangle^1$ and meets a student of Alice in state $|s_1\rangle^A$, three students of qubit 1 which were previously in the relative world $|1\rangle^1$ now enter relative world $|s_1\rangle^1$ and meet students of Alice in state $|s_1\rangle^A$, and one student of qubit 1 which was previously in the relative world $|1\rangle^1$ now enters relative world $|s_2\rangle^1$ and meets a student of Alice in state $|s_2\rangle^A$.

For any setting, the referee then writes the outcomes in the external memory of the students of both Alice and qubit 1, and the state $|\psi^{1,2,A} \rangle =  U^{1,A}|\psi_0\rangle^{1,2}|R\rangle^A$ in their internal memories.  To save on students during the measurement interaction, the same 8 students playing lives of qubit 1 now also play lives of Alice, and after the interaction they continue as lives of Alice.

The situation is symmetrically identical for Bob, so after the measurement, his 8 students will now have $|\psi^{1,2,B}\rangle = U^{2,B}|\psi_0\rangle^{1,2}|R\rangle^B$ written in their internal memories.

Now Alice's 8 students and Bob's 8 students reunite.  When two systems meet, their internal memories synchronize by accumulating all states and coupling unitaries from both of their past internal memories.  When students of Alice and Bob meet, their internal memories synchronize to $|\psi\rangle^{1,2,A,B} = U^{1,A} U^{2,B}|\psi_0\rangle^{1,2}|R\rangle^A |R\rangle^B$, and their preferred bases remain the same, so the proportion of students in each relative world who meet is given by $P(s_a, s_b, h) = |\langle \psi|^{1,2,A,B} | s_a\rangle^A | s_b\rangle^B |h\rangle^{1,2}|^2 = |\langle \psi_0|^{1,2} | s_a\rangle^1 | s_b\rangle^2 |^2$  , where $s_a$ is the $a$th outcome of Alice's measurement, and $s_b$ is the $b$th outcome of the Bob's measurement, and $|h\rangle^{1,2}$ is the most recent external memory state of qubits 1 and 2 within the history $h$.  The proportion of lives that meet with each pair of specific histories is determined using Eq. \ref{Phistory}.

If Alice and Bob measured the same setting, then four students of Alice who got $|s_1\rangle^A$ each meet a student of Bob who got $|s_2\rangle^B$, and four students of Alice who got $|s_2\rangle^A$ each meet a student of Bob who got $|s_1\rangle^B$.  If they measured different settings, then three of Alice's students who got $|s_1\rangle^A$ each meet a student of Bob who got $|s_1\rangle^B$, one of Alice's students who got $|s_1\rangle^A$ meets a student of Bob who got $|s_2\rangle^B$, three of Alice's students who got $|s_2\rangle^A$ each meet a student of Bob who got $|s_2\rangle^B$, and one of Alice's students who got $|s_2\rangle^A$ meets a student of Bob who got $|s_1\rangle^B$.  In any case, it is easy to see that the entanglement correlations of the initial state $|\psi_0\rangle^{1,2}$ have been obeyed for the entire group of students.

Then, by repeating the exercise many times, an individual student will experience Born rule statistics and thus a violation of a Bell inequality, even though everything was done obeying explicit local causality.

\end{document}